\documentclass{article}
\usepackage[utf8]{inputenc}

\newcommand{\doctitle}{Inference of phase field fracture models}
\newcommand{\docheader}{\doctitle}
\newcommand{\pdftitle}{\doctitle}

\newcommand{\docauthor}{}

\newcommand{\docfooter}{University of Michigan}

\usepackage{main}	
\begin{document}

\title{\Large\doctitle}

\author[1,2]{Elizabeth Livingston} 
\author[1,2]{Siddhartha Srivastava}
\author[2,3]{Jamie Holber} 
\author[5]{Hashem M. Mourad}
\author[1,2,3,4]{Krishna Garikipati\thanks{Corresponding author at: Department of Mechanical Engineering, University of Michigan, United States. \emph{E-mail address}: krishna@umich.edu (K. Garikipati).}} 
\affil[1]{Department of Mechanical Engineering, University of Michigan, United States}
\affil[2]{Michigan Institute for Computational Discovery \& Engineering, University of Michigan, United States}
\affil[3]{Applied Physics, University of Michigan, United States}
\affil[4]{Department of Mathematics, University of Michigan, United States}
\affil[5]{T-3, Theoretical Division, Los Alamos National Laboratory, United States}
\date{} 

\renewcommand{\baselinestretch}{1} 
\maketitle


\begin{abstract}
\pagestyle{abstract}

The phase field approach to modeling fracture uses a diffuse damage field to represent a crack. This addresses the singularities that arise at the crack tip in computations with sharp interface models, mollifying some of the difficulties associated with the mathematical and numerical treatment of fracture. The introduction of the diffuse field helps with crack propagation dynamics, enabling phase-field approaches to model all phases of damage from crack initiation to propagation, branching, and merging. Specific formulations, beginning with brittle fracture, have also been shown to converge to classical solutions. 
Extensions to cover the range of material failure, including ductile and cohesive fracture, leads to an array of possible models.
There exists a large body of studies of these models and their consequences for crack  evolution. However, there have not been systematic studies into how optimal models may be chosen. Here we take a first step in this direction by developing formal methods for identification of the best parsimonious model of phase field fracture given full-field data on the damage and deformation fields. We consider some of the main models that have been used to model damage, its degradation of elastic response, and its propagation. Our approach builds upon Variational System Identification (VSI), a weak form variant of the Sparse Identification of Nonlinear Dynamics (SINDy). In this first communication we focus on synthetically generated data but we also consider central issues associated with the use of experimental full-field data, such as data sparsity and noise.

\end{abstract}


\newpage
\clearpage
\singlespacing
\pagenumbering{arabic}
\pagestyle{document}


\section{Introduction}

\subsection{Phase field modeling of fracture}


In the phase field method, a sharp, discontinuous crack is approximated using a diffuse, continuous damage field $d$. This field represents the state of the material (undamaged $d=0$ and damaged $d=1$) which evolves in response to external loading conditions. The total energy, $\Psi(\boldsymbol{u},d)$, is minimized with respect to the displacement $\boldsymbol{u}$ and internal state $d$ resulting in a quasi-static equilibrium. Viscous regularization of  rate-independent damage can be written  as a gradient flow, with the variational derivative of the free energy driving the evolution of the damage field. The phase-field approach can capture all stages of complex fracture patterns from crack nucleation, to propagation, merging and branching in a single model without including or resolving discontinuities in the solution fields. Over recent decades, the phase field fracture approach has gained significant attention in the computational mechanics community due to its effectiveness in capturing complex fracture patterns (e.g., crack merging and branching), which have proved challenging to finite element methods based on discontinuity-resolving formulations.\cite{rudraraju2012predictions}

The phase-field approximation was first introduced in the mechanics community by Francfort and Marigo, extending Griffith's theory for brittle fracture by using energy minimization of the variational model.\cite{francfort_revisiting_1998,griffith_phenomena_1920} As in Griffith's theory of brittle fracture, the total energy functional is the sum of the bulk energy in the material and the crack surface energy, and competition between these terms drives the crack. Initial numerical implementations by Bourdin, Francfort, and Marigo introduced a regularized formulation for the crack surface energy that utilized the smooth damage field $d$, allowing for a continuous displacement field $\boldsymbol{u}$.\cite{bourdin_numerical_2000} This form was inspired by work from Ambrosio and Tortorelli which proved the $\Gamma$-convergence of the minimization of a two-field, regularized, elliptic functional.\cite{ambrosio_approximation_1990} Despite the $\Gamma$-convergence of the regularized form, unrealistic crack patterns could appear in compression. Amor, Marigo, and Maurini split the elastic energy of the material into volumetric and deviatoric components with damage only degrading the  part of the elastic energy depending on the tensile volumetric strain.\cite{amor_regularized_2009} This work also introduced a length scale parameter $\ell$ representing the size of the transition layer between damaged and undamaged states. To handle irreversibility of a crack in the damage evolution equation, i.e. non-healing cracks, Miehe et al. introduced a history variable that represents the maximum of the tensile volumetric strain-dependent elastic energy over time.\cite{miehe_phase_2010} This variable allowed for the algorithmic decoupling of the governing equations enabling the use of a staggered solution scheme. The governing equations were derived using thermodynamic and continuum mechanics arguments and a  split of the elastic energy using the spectral decomposition of the strain. 

A similar development of models exists within the physics community, based on the Landau-Ginzburg phase transition. Motivated by this body of work, Kuhn and Muller, reformulated the approach using a system of stress-equilibrium equations.\cite{kuhn_phase_2008} Similarly, Miehe et al. introduced the \textit{crack driving force} governed by the \textit{crack driving state function} in the damage evolution equation.\cite{miehe_phase_2015} This function was represented using both strain and stress criteria replacing the elastic energy. A comprehensive overview of the development of models in the mechanics and physics communities may be found in the review from Ambati, Gerasimov, and de Lorenzis.\cite{ambati_review_2015}

Beyond the key developments listed above, there have been extensions as well as alternate approaches for different types of fracture. While the phase-field method was developed based on the theory of brittle fracture, the choice of degradation function affecting the bulk energy in the material is crucial in modeling the relationship between the damage field and elastic properties. In a brittle material, the response is typically linear elastic up until failure, however, the quadratic degradation function used in the initial literature mentioned above leads to a significant degradation in stiffness. Many alternatives have been considered in the literature, and discussions may be found in Kuhn, Schluter, and Muller and Sargado et al.\cite{kuhn_degradation_2015,sargado_highaccuracy_2018}

The representation of the crack surface energy has also taken a few different forms in the literature. To include linear elastic behavior before the onset of damage, Pham et al. considered a linear function of damage ($d$) instead of the typical quadratic form ($d^2$) in the regularized crack surface energy.\cite{pham_gradient_2011} In the literature, these are referred to as the "AT1" and "AT2" models respectively, based on the work of Ambrosio and Tortorelli.\cite{ambrosio_approximation_1990} Non-monotonic functions may also be considered, such as the unified brittle/cohesive model of Wu et al., $2d-d^2$, and a double-well function ($d^2(1-d)^2$) often used in the physics community.\cite{wu_unified_2017} The variations described here along with others for ductile, cohesive, and dynamic fracture have been  summarized in review papers by Bourdin, Wu, and Diehl.\cite{bourdin_variational_2008,wu_phasefield_2020, diehl_comparative_2022}

Attempts to unify the variety of phase-field models for different fracture modes,\cite{freddi_regularized_2010} brittle and quasi-brittle fracture,\cite{wu_unified_2017} and cohesive fracture\cite{suh_phase_2020} include more generalized forms of the functions that govern crack surface energy and elastic energy degradation. Studies by Kuhn, Ambati, Alessi, and Svolos  compared the mathematical and numerical performance of various formulations and some suggest a general formulation.\cite{kuhn_degradation_2015,ambati_review_2015,alessi_comparison_2018,svolos_convexity_2023} The general formulation introduced by Svolos et al., includes a set of parameters and functions that vary between formulations.\cite{svolos_convexity_2023}
Using a similar formulation, non-exhaustive lists are included in \cref{sec:phase_field_fracture_theory} for the degradation functions (\ref{tab:degradation_functions}), smooth fracture energy functions (\ref{tab:geometric_functions}), regularization parameters (\ref{tab:crack_energy_models}), and history update approaches (\ref{tab:history_functions}). 
The different approaches are often selected and tuned by minimizing the difference between the numerical estimation of the material state (damage) and the experimental results showing crack growth. Finally, we note phase field fracture models are remarkably similar to gradient-enhanced continuum damage models.\cite{deborst_gradient_2016} Their numerical treatment has led to the development of a range of finite element methods.\cite{wells2004discontinuous,molari2006discontinuous}
\subsection{Variational system identification}

The construction of mathematical models using observed data is fundamental in science and engineering. Initially, models were constructed in order to fit the observed data. In recent years, the abundance of data, advances in mathematical methods and improvements in computational efficiency have enabled the development of data-driven methods of solving inverse problems, identifying and inferring models with fewer assumptions. One traditional approach to solving an inverse problem is to estimate the parameters of a chosen model. This approach is common for identifying  material parameters as well as coefficients of mathematical terms such as those in various functions in the phase field approach to fracture. We will refer to it as \emph{parameter identification}. However, beyond fitting the parameters of a chosen form, the terms of a model themselves can be chosen to optimally represent the available data. In this approach, the candidate terms or operators are combined in a cost function and optimization techniques are used to identify the the most important terms while imposing some form of model parsimony. We will refer to this class of approaches as \emph{system identification}. It is the focus of the current work.

In recent work, both the material parameters and coefficients of mathematical terms in various functions of phase field fracture models have been determined using parameter identification. In Wu et al. \cite{wu_parameter_2021}, a Bayesian approach was used to estimate the material parameters: bulk and shear moduli, tensile strength, as well as the fracture toughness of a brittle cement mortar. Data was limited to the force-displacement curve from a single three-point bending experiment. Therefore, the linear elastic regime was used to identify the parameters, which were then validated in the inelastic regime. The elasticity parameters may be identified from the elastic regime, and the yield strength from the end of the elastic regime. However, the fracture toughness is directly related to the initiation and propagation of damage in the inelastic regime and therefore has a complex dependence on the terms in the selected phase field fracture model. These terms also  have a similarly complex influence on the force-displacement curve, since the latter is a highly integrated, reduced dimensional representation of the physics of crack initiation and progression. On the other hand, full-field data, if available, may show patterns in damage fields that are precursors to crack initiation and  propagation. Such information is typically lost in the spatial integration that is performed for a load-displacement curve. Therefore, full-field data-driven approaches could lead to more detailed and possibly precise system identification of phase field fracture models.

In Kosin et al.,\cite{kosin_parameter_2024} integrated digital image correlation was used to estimate the boundary conditions, Poisson ratio, fracture energy, and internal length. Full spatio-temporal synthetic data was utilized to infer these parameters. However, the data was masked to avoid integration errors near the crack. Synthetic data allowed for the analysis of the reduction in the residual and parameter accuracy since the exact solution was known. These studies focused on the estimation of material and damage parameters while assuming a specific formulation for the phase-field model. This approach to parameter estimation thus relied on pre-existing knowledge, or assumptions, of the material behavior and mechanics of damage propagation. Data-driven system identification, however, may also be used to to infer the actual mechanisms of fracture by identifying the governing equations with fewer assumptions about the underlying physics. 

The data-driven approach and term ``system identification'' has roots in the dynamics and controls community. Lennart Ljung outlined the standard procedure for system identification: (1) data observation, (2) set of candidate models, (3) criterion of fit, and (4) model validation.\cite{lennart_ljung_system_1999} Schmidt and Lipson  extended this idea to identify "free-form natural laws" with minimal prior knowledge of the model form  using symbolic regression.\cite{schmidt_distilling_2009} In order to avoid nontrivial solutions, the algorithm was designed to connect groups of variables. The best candidate analytical expressions were determined by maximizing accuracy (agreement with the data) and parsimony (simplicity of the model) on a Pareto front. To reduce the computational expense and scale to larger dynamical systems, Brunton et al. leveraged compressive sensing and sparse regression.\cite{brunton_discovering_2016} The sparse identification of nonlinear dynamics (SINDy), is limited by the choice of basis functions and measurement coordinates that must enable sparse representation.

SINDy is similar to dynamic mode decomposition, which can be seen as an equation-free method that identifies Koopman modes to fit linear operators to a nonlinear dynamical system.\cite{rowley_spectral_2009} Similarly, a neural network may be used to map input data to an output without analytical expressions.\cite{gonzalez-garcia_identification_1998} Approaches to introduce the physics into a neural network include structures like the Physics Informed Neural Networks (PINNs) from Raissi and Variational Onsager Neural Networks (VONN) from Huang.\cite{raissi_physics-informed_2019, huang_variational_2022}  Others, like the Constitutive Artificial Neural Network (CANN) from Linka, learn a constitutive model for materials by using neural networks with a library of  activation functions.\cite{linka_new_2023}

While the various forms of neural network architectures were explored and system identification was applied to dynamical systems in the strong form, the approach of discovering governing equations using regression was extended to the weak form by Wang et al.\cite{wang_variational_2019} The authors referred to their approach as Variational System Identification. In the variational setting, the Neumann boundary conditions may be identified in addition to the differential and algebraic terms within the weak form of the partial differential equation. There is also a reduced requirement for smoothness of the solution evaluated using the data since  derivatives are transferred to the weighting functions by carrying out integration by parts. Finally,  step-wise regression was used to identify the best candidate model--although sparsification could also be achieved by the $\mathrm{L}^1$ norm (the LASSO approach). The approach of Variational System Identification has since been extended to the Bayesian setting \cite{wang_perspective_2020}, as well as applied to sparse and incomplete data in materials,\cite{wang_variational_2021} soft material constitutive modeling,\cite{wang_inference_2021,nikolov2022ogden} dynamics of cell migration,\cite{ho2023oscillatory,srivastava_pattern_2024,kinnunen2024inference} and the epidemiology of COVID.\cite{wang2020system, wang2021system} Similar approaches utilizing the weak form of a PDE were explored by Messenger et al. and Wentz et al. \cite{messenger_weak_2021,wentz_derivative-based_2023} which both utilized sparse regression. Since the phase-field fracture formulation is of variational  origin, the weak form is a natural setting for system identification. Here we adopt Variational System Identification as described in the collection of literature that has been cited above, especially in the context of material constitutive modeling.\cite{wang_inference_2021,nikolov2022ogden}

Specifically, we use regression to identify the form of the hyperelastic free energy density function and damage growth laws that best represent the data from a set of admissible candidates.   We begin this work with a review of the existing phase-field fracture models and solution approaches ( \cref{sec:phase_field_fracture_theory}--\cref{sec:fwdsoln}), motivating the need for a system identification framework (\cref{sec:inference}). Using computationally generated, high-fidelity, spatio-temporal data (\cref{sec:datagen}) this approach is then validated on a sample problem (\cref{sec:inferenceresults}). We also consider the effect of noise and sub-sampling on the robustness with which the models are identified. Concluding remarks are presented in \cref{sec:discussion}.

\section{Phase field fracture theory}
\label{sec:phase_field_fracture_theory}

\begin{figure}[ht!]
    \centering
    \includegraphics[width=0.5\linewidth]{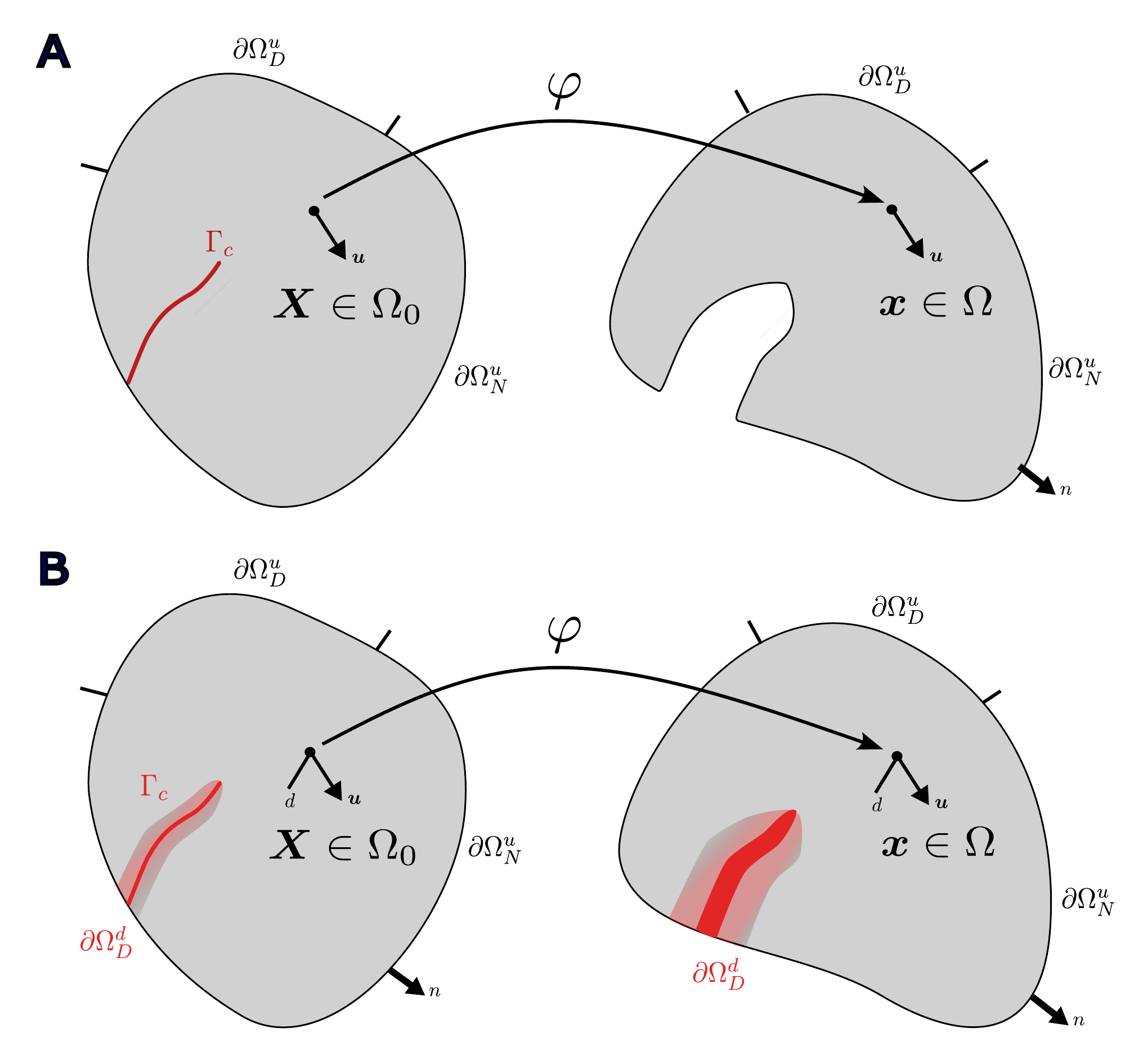}
    \caption{Fields of interest and associated boundary conditions: (a) typical viewpoint of a discrete crack $\Gamma_c$ within a deformable body (b) regularization of the crack surface via the introduction of a phase field $d$. }
    \label{fig:pff_potato}
\end{figure}

In classical approaches to fracture, a sharp crack set, $\Gamma_c\subset\mathbb{R}^{n-1}$, is represented as a discontinuity within a solid body $\Omega\subset\mathbb{R}^n$. The displacement field is $\boldsymbol{u}:\Omega\rightarrow \mathbb{R}^n$, $\boldsymbol{u}(\boldsymbol{X},t)\in BD(\Omega)=\{\vec{v}|\vec{v}=U \textrm{on}\ \partial\Omega_D^u\}$ where $BD(\Omega)$ is the space of bounded deformations for which all components of the strain are bounded measures. The boundary $\partial\Omega$, with outward normal $\boldsymbol{n}$, is split into two disjoint subsets $\partial\Omega_D^u$, and $\partial\Omega_N^u$ for the application of Dirichlet and  Neumann boundary conditions, respectively as shown in \ref{fig:pff_potato}(a). This figure also shows the smooth deformation of the continuum body in the reference configuration with material points $\boldsymbol{X}\in\Omega_0$ to the current configuration $\boldsymbol{x}\in\Omega$, $\boldsymbol{x} = \boldsymbol{\varphi}(\boldsymbol{X})$. This is a one-to-one mapping as shown in \ref{eq:deformation_map}. 

\begin{equation}
    \varphi:\Omega_0 \rightarrow \Omega \subset\mathbb{R}^n
    \label{eq:deformation_map}
\end{equation}

In the phase-field approach to fracture, the crack set, $\Gamma_c$, is represented using a diffuse or ``smeared out'' scalar phase-field $d(\boldsymbol{x},t)$ which is equal to 1 in the crack and 0 in the undamaged material. The evolution of $d$ is governed by a differential equation that is distinct from the governing equations of elasticity, requiring independent Dirichlet boundary conditions $\partial\Omega_D^d$ as shown in \ref{fig:pff_potato}(b). Due to the regularization introduced by the damage field in this method, the displacement, $\boldsymbol{u}:\Omega\rightarrow \mathbb{R}^n$, belongs to the Sobolev space of square integrable functions, $\boldsymbol{u}(\boldsymbol{x},t)\in\boldsymbol{H}^1(\Omega)$.

Using this displacement field, the deformation gradient is written as

\begin{equation}
    \boldsymbol{F} = \pdv{\varphi(\boldsymbol{X})}{\boldsymbol{X}} = \boldsymbol{I} + \nabla \boldsymbol{u}
    \label{eq:deformation_gradient}
\end{equation}

and defines the right Cauchy-Green tensor

\begin{equation}
    \boldsymbol{C} = \boldsymbol{F}^T \boldsymbol{F}.
    \label{eq:right_cauchy_green}
\end{equation}

The free energy density used to represent materials with fracture is inspired by Griffith's theory of brittle fracture\cite{griffith_phenomena_1920} which accounts for the bulk elastic energy in the bulk and the surface energy of the crack: \ref{eq:energy_density}.
\begin{equation}
    \psi(\boldsymbol{C}, d) = \underbrace{g(d)  \psi_e(\boldsymbol{C})}_{\text{bulk energy in damaged material}} + \underbrace{g_c \gamma_\ell(d, \nabla d)}_{\text{surface energy of crack}}
    \label{eq:energy_density}
\end{equation}
The first term represents the decay of the elastic energy factored into the degradation function, $g(d)$ and the elastic energy of the undamaged material, $\psi_e(\boldsymbol{C})$.  In the second term, $g_c$ is a material specific constant known as the critical energy release rate or fracture toughness and $\gamma_\ell(d,\nabla d)$ is the regularized fracture energy. The components of the free energy density are described in the following section along with  the solution procedure used in this work.

\subsection{Energy of the damaged material}

\subsubsection{Elastic energy}
\label{sec:elastic_energy}

Consider first a linear isotropic solid, with the elastic strain energy density

\begin{equation}
    \psi_0(\boldsymbol{\varepsilon}) = \lambda(\varepsilon_1+\varepsilon_2+\varepsilon_3)^2/2 + \mu(\varepsilon_1^2+\varepsilon_2^2+\varepsilon_3^2)
    \label{eq:strain_energy_linear}
\end{equation}

as a function of the infinitesimal strain tensor $\boldsymbol{\varepsilon}$ and Lam\'e parameters $\lambda$ and $\mu$. The components in this form are the principal strains $\{\varepsilon_a\}_{a=1,2,3}$ and principal strain directions $\{n_a\}_{a=1,2,3}$, which give the strain tensor $\boldsymbol{\varepsilon} = \sum_{a=1}^3 \varepsilon_a n_a\otimes n_a$. 

Using the full strain tensor when evaluating the elastic energy allows for unrealistic crack initiation and propagation in compression. Decomposing the elastic energy into compressive ($\psi_0^-$) and tensile ($\psi_0^+$) components, we can apply the degradation function to the latter, eliminating damage-induced degradation of the compressive component of strain energy density.

\begin{equation}
    \psi(\boldsymbol{\varepsilon},d) = g(d)\psi_0^+(\boldsymbol{\varepsilon}) + \psi_0^-(\boldsymbol{\varepsilon})
    \label{eq:energy_decomposition}
\end{equation}

In this work, the energy is split via the spectral decomposition, as adopted by Meihe et al. \cite{miehe_phase_2010}. Using the linear isotropic elasticity from \ref{eq:strain_energy_linear}, the positive and negative components are defined in \ref{eq:energy_compressive_tensile_split}.

\begin{equation}
    \psi_0^\pm(\boldsymbol{\varepsilon}) = \lambda\langle\varepsilon_1+\varepsilon_2+\varepsilon_3\rangle_\pm^2/2 + \mu(\langle\varepsilon_1\rangle_\pm^2 + \langle\varepsilon_2\rangle_\pm^2 + \langle\varepsilon_3\rangle_\pm^2)
    \label{eq:energy_compressive_tensile_split}
\end{equation}

The Macaulay bracket operation is $\langle*\rangle_\pm = (\abs{*} \pm *)/2$, where the ``negative'' Macaulay bracket rectifies negative values of the input $*$ to positive values (thus working opposite the ``positive'' Macaulay bracket). 
A comparison of alternate approaches for splitting the energy may be found in Li et al. \cite{li_gradient_2016} and a stress based decomposition may be found in Steinke and Kaliske.\cite{steinke_phasefield_2019}

In order to extend beyond small strain, while maintaining the energy split for damage, we consider the Saint-Venant Kirchoff model, \ref{eq:saint_venant_kirchoff}. This model includes nonlinearity in the displacement gradients while maintaining a linear stress-strain relationship by using the Green-Lagrange strain tensor $\boldsymbol{E}=\frac{1}{2}(\boldsymbol{F}^T\boldsymbol{F}-\boldsymbol{I})$:
\begin{equation}
    \psi_e = \frac{1}{2}\lambda(\trace{\boldsymbol{E}})^2 + \mu(\trace{\boldsymbol{E}^2})
    \label{eq:saint_venant_kirchoff}
\end{equation}

\subsubsection{Degradation function}

The degradation function $g(d)$ in the first term of \ref{eq:energy_density} is continuous and has the  properties:
\begin{equation}
   g(0) = 1, \hspace{2em} g(1) = 0, \hspace{2em} g'(1) = 0
\label{eq:degradation_function_properties}
\end{equation}

\begin{figure}[ht!]
    \centering
    \includegraphics[width=0.8\linewidth]{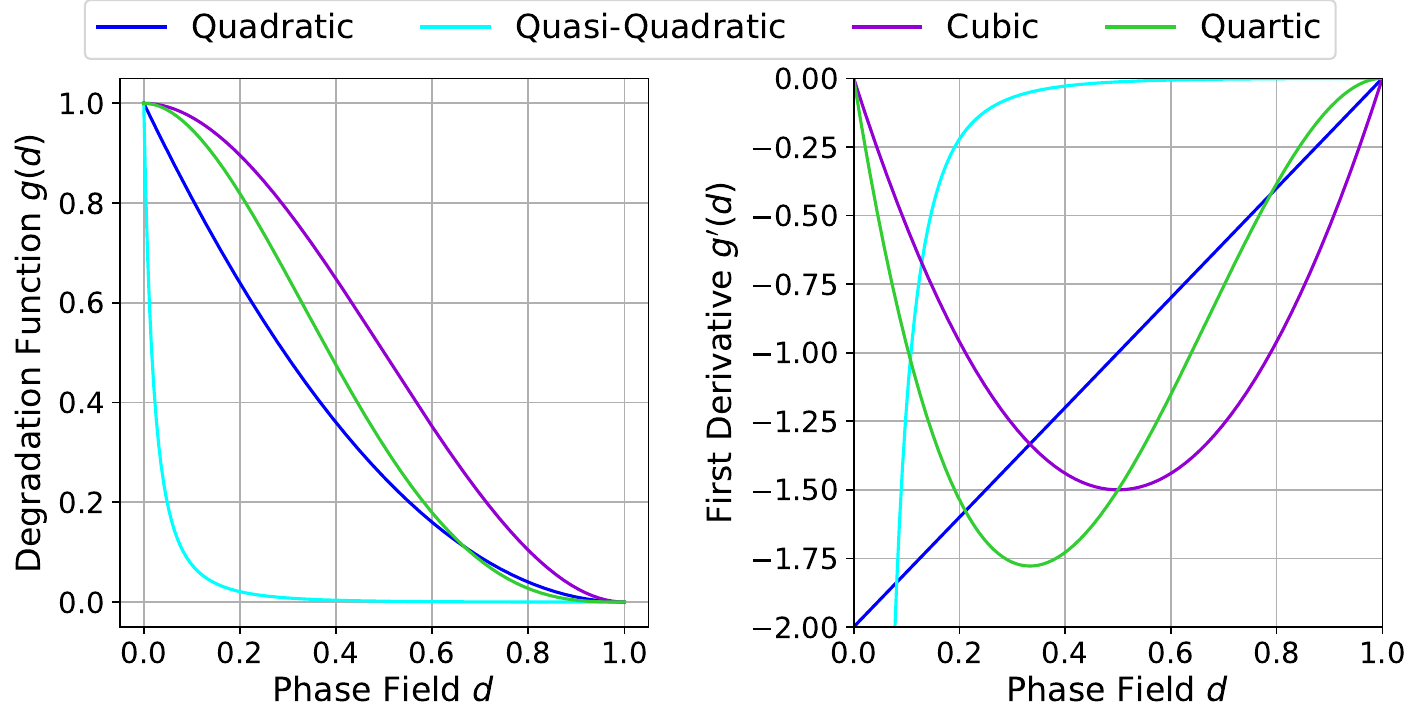}
    \caption{Common degradation functions from the literature and their first derivatives. The exact form of the plotted functions is shown in \ref{tab:degradation_functions}.}
    \label{fig:degradation_functions}
\end{figure}

\vspace{-1em} The first two conditions, respectively, ensure that the elastic energy is fully intact (i.e. there is no degradation) when there is no damage ($d=0$) and that the elastic energy is vanishes if the material is fully damaged ($d=1$). The vanishing gradient of the fully damaged material in the third condition is relevant under quasi-static equilibrium, and is discussed in \cref{sec:damage_evolution}. It imposes continuous differentiability and  eliminates the crack driving term in a fully damaged material. There are a number of functions that may be used while satisfying these properties. A summary, reproduced from Svolos et al.,\cite{svolos_convexity_2023} of some degradation functions used in the literature, and that satisfy these conditions is presented in \ref{tab:degradation_functions} with illustrations of the functions and their derivatives in \ref{fig:degradation_functions}.

\begin{table}[!ht]
\centering
\begin{tabular}{l l l p{0.3\linewidth}} \hline
Abbr & Type  & $g(d)$   & References\\ \hline
(G1) & Quadratic  & $(1-d)^2$   & Bourdin \cite{bourdin_numerical_2000}, Miehe \cite{miehe_phase_2010} \\
(G2) & Cubic  & $(3+s_0)(1-d)^2-(2+s_0)(1-d)^3$ & $s_0=0$ Kuhn \cite{kuhn_degradation_2015}, $s_0\approx 0^-$, Borden \cite{borden_phasefield_2016}  \\
(G3) & Quartic  & $4(1-d)^3-3(1-d)^4$  & Karma \cite{karma_phasefield_2001} \\
(G4) & Power Function & $(1-d)^p$  & Pham \cite{pham_gradient_2011} \\
(G5) & Polynomial     & $(p+s_0)(1-d)^{p-1}-(p-1+s_0)(1-d)^p$  & $s_0\in [-p,0]$ \\ 
(G6) & Quasi-Linear   & $\frac{1-d}{1-d+\alpha_1 d}$  & $\alpha_1=-s_0$ Geelen \cite{geelen_phasefield_2019} \\
(G7) & Quasi-Quadratic & $\frac{(1-d)^2}{(1-d)^2+\alpha_1 d+\alpha_1\alpha_2 d^2}$ & $\alpha_1=-s_0,\alpha_2\ge 1$ Lorentz \cite{lorentz_modelling_2012}, $\alpha_1=-s_0,\alpha_2=-\frac{1}{2}$ Alessi \cite{alessi_comparison_2018} \\
(G8) & Rational & $\frac{(1-d)^p}{(1-d)^p+Q_p(d)}$  & $p\ge 3,\alpha_1=-s_0$ Wu \cite{wu_phasefield_2020} \\ \hline
\end{tabular}
\caption{Degradation functions and references from L. Svolos et al. \cite{svolos_convexity_2023}}
\label{tab:degradation_functions}
\end{table}

\subsection{Regularized fracture energy}
\label{sec:regularize_fracture_energy}

The regularized fracture energy density, $\gamma_\ell$ in \ref{eq:energy_density} has been proposed in several forms in the literature. The generalized form in \ref{eq:crack_energy_generalized}. has been written with  coefficient of $1$ on the gradient term, $\frac{1}{2}|\nabla d|^2$, which will be useful when setting up the inference problem in \cref{sec:inference}. Beyond the coefficients, the key component governing the damage evolution within the regularized fracture energy is the smooth fracture energy, $f(d)$. While this term does not distinguish between a uniform field $d$ and one that is smooth but has high gradients, the term $\vert\nabla d\vert^2$ imposes penalization, increasing the regularized fracture energy at the crack, where the gradient in damage is large.  The inclusion of a gradient term thus enforces a diffuse crack representation via the field $d$. The competition between the elastic free energy density and these fracture energy terms enforces continuous fields for the displacement $\boldsymbol{u}$, and strain $\boldsymbol{E}$.

\begin{equation}
    \gamma_\ell(d,\nabla d) = c_s \left(\beta f(d) + \frac{1}{2} |\nabla d|^2 \right)
    \label{eq:crack_energy_generalized}
\end{equation}

\subsubsection{Smooth fracture energy}
\label{sec:smooth_fracture_function}

The smooth fracture energy function, $f(d),$ controls the dependence on  damage $d$ without regard to how the local neighborhood is positioned relative to a crack. As discussed in the introduction, early implementations of the regularized fracture energy used a quadratic function, $f_2(d)=d^2$, for the smooth fracture energy. While its convexity improves the stability of the overall formulation, damage initiates as soon as a material is loaded. A linear function, $f_1(d)=d$, allows for an elastic region before the onset of damage. In the literature, these are often referred to as the \texttt{AT2} and \texttt{AT1} models respectively, based on the work of Ambrosio and Tortorelli.\cite{ambrosio_approximation_1990} In the physics community, a double well function, $f_3(d) = d^2(1-d)^2$, is considered because it provides an energy barrier at a local maximum between undamaged and fully damaged states or ``phases''.\cite{karma_phasefield_2001} However, it has been observed that the existence of two energetically equivalent local minima for the  undamaged and damaged states, which can lead to nonphysical crack propagation.\cite{wu_unified_2017,kuhn_degradation_2015} These functions are summarized in \ref{tab:geometric_functions} and illustrated in \ref{fig:geometric_functions}.

\begin{minipage}{\textwidth}
    \begin{minipage}[m]{0.45\textwidth}
    \centering
        \begin{tabular}{ll} \hline
        Type            & $f(d)$                          \\ \hline
        Linear          & $f_1(d) = d$               \\
        Single Well     & $f_2(d) = d^2$             \\
        Double Well     & $f_3(d) = d^2(1-d)^2$     \\ \hline
        \end{tabular}
        \captionof{table}{Smooth fracture energy functions used in the literature.}
        \label{tab:geometric_functions}
    \end{minipage}
    \hfill
    \begin{minipage}[m]{0.49\textwidth}
            \centering
            \includegraphics[width=0.8\linewidth]{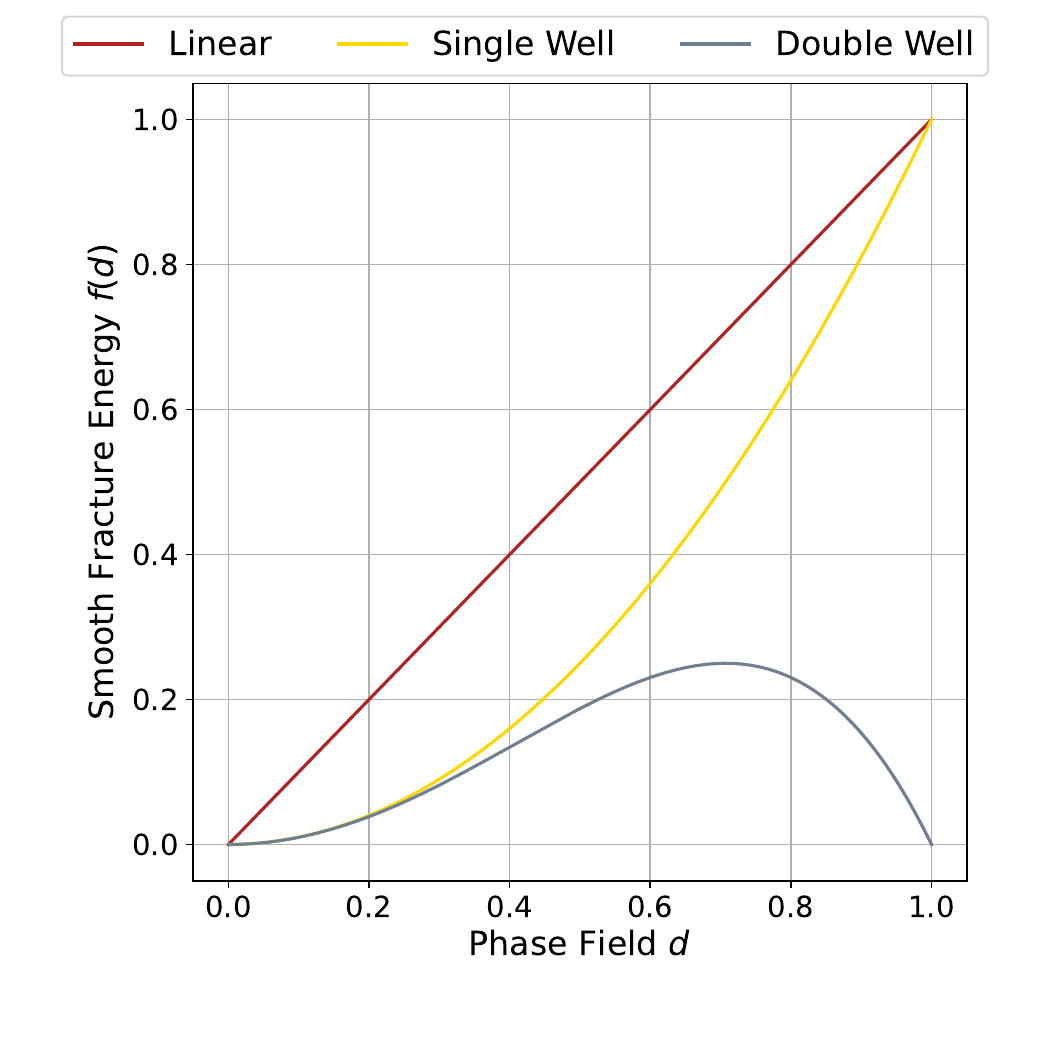}
            \captionof{figure}{Visualization of the three primary smooth fracture energy functions, $f(d)$, used in the literature: Linear $f_1(d)=d$, Single Well $f_2(d)=d^2$, and Double Well $f_3(d)=d^2(1-d)^2$.}
            \label{fig:geometric_functions}
  \end{minipage}
\end{minipage}

\subsubsection{Coefficients and crack length parameter}

The coefficients $c_s$ and $\beta$ determine the relationship between the gradient term, $\frac{1}{2}|\nabla d|^2$, and the smooth fracture energy function $f(d)$. In order to govern the size of the diffuse region of damage, these coefficients are often defined as a function of the crack length-scale $\ell$, which determines the width of the diffuse crack in the regularized domain. In numerical implementations, this parameter has a lower bound since it must be greater than the size of a single element. While this may be considered a numerical parameter for regularization of the sharp crack, it admits interpretation as a material parameter because it controls the transition between undamaged and fully damaged regions at the fracture surface.\cite{amor_regularized_2009} Forms of regularized fracture energy used in the literature are summarized in \ref{tab:crack_energy_models} with the coefficients $c_s$ and $\beta$ of \ref{eq:crack_energy_generalized}. This table follows Svolos et al.\cite{svolos_convexity_2023} which uses a similar generalized form.

\begin{table}[!ht]
\centering
\caption{Regularized fracture energy formulations and associated references. A modified version of Table 3 following Svolos et al.\cite{svolos_convexity_2023}}
\begin{tabular}{c|c|c|c}
\hline Abbr. & $c_s$ & $\beta$ & Reference \\ \hline
(P1) & $2\ell$ & $\frac{1}{8\ell^2}$ & Bourdin et al. \cite{bourdin_numerical_2000} Borden et al. \cite{borden_phasefield_2016}  \\
(P2) & $2$ & $\frac{1}{4}$ & Karma et al. \cite{karma_phasefield_2001} \\
(P3) & $\ell$ & $\frac{1}{2\ell^2}$ & Miehe et al. \cite{miehe_phase_2010}, Miehe and Mauthe \cite{miehe_phase_2016} \\
(P4) & $\frac{2\ell}{\pi}$ & $\frac{1}{2\ell^2}$ & Alessi et al. \cite{alessi_gradient_2015}, Wu \cite{wu_unified_2017}  \\
(P5) & $\frac{3\ell}{4}$ & $\frac{1}{2\ell^2}$ & Geelen et al. \cite{geelen_phasefield_2019} \\
(P6) & $\frac{3\ell}{4\sqrt{2}}$ & $\frac{1}{\ell^2}$ & Pham et al. \cite{pham_gradient_2011}
\end{tabular}
\label{tab:crack_energy_models}
\end{table}

\subsection{Damage evolution}
\label{sec:damage_evolution}

In the absence of external body forces and tractions, the total free energy of the system is:
\begin{equation}
    \Psi(\boldsymbol{u},d) = \int_{\Omega} g(d)\psi_e(\boldsymbol{C},d) dV + \int_{\Omega} g_c \gamma_\ell(d,\nabla d) dV
    \label{eq:total_energy}
\end{equation}

The governing equations are obtained by different approaches to free energy minimization. For this we consider the first variations of the functional $\Psi$:  $\delta_{\boldsymbol{u}}\Psi$ and $\delta_d \Psi$. The resulting variational derivatives appear in \ref{eq:euler_lagrange-elasticity} and \ref{eq:euler_lagrange-damage}. 

\begin{align}
        \delta_{\boldsymbol{u}}\Psi(\boldsymbol{u},d) &= \int_\Omega g(d) \frac{\partial\psi_e(\boldsymbol{C},d)}{\partial\boldsymbol{F}}:\nabla \boldsymbol{w}_{\boldsymbol{u}} \mathrm{d}V, 
        \label{eq:euler_lagrange-elasticity} \\
    \delta_d\Psi(\boldsymbol{u},d) &= \int_\Omega g'(d)w_d \psi_e(\boldsymbol{C},d) \ \mathrm{d}V + \int_\Omega g_c c_s \left( \beta f'(d)w_d + \nabla w_d \cdot \nabla d \ \right) \mathrm{d}V,
        \label{eq:euler_lagrange-damage}
\end{align}

recalling that $\boldsymbol{C}$ and $\boldsymbol{F}$ involve $\nabla\boldsymbol{u}$. A complete derivation appears in  \cref{appendix:euler_lagrange}.

The variations $\boldsymbol{w}_{\boldsymbol{u}} \in V_{\boldsymbol{u}}$ and $w_d \in V_d$ are defined as:

\begin{align*}
    V_{\boldsymbol{u}}&=\{\boldsymbol{w}_{\boldsymbol{u}}:\Omega\rightarrow\mathbb{R}^n \ \vert \boldsymbol{w}_{\boldsymbol{u}} \in [H^1(\Omega)]^n, \hspace{0.8em} \boldsymbol{w}_{\boldsymbol{u}}\vert_{\partial \Omega^{\boldsymbol{u}}_D} = \boldsymbol{0}\} \\
    V_d&=\{w_d:\Omega\rightarrow\mathbb{R} \hspace{0.9em} \vert w_d \in [H^1(\Omega)]^{n-1},w_d\vert_{\partial\Omega^d_D}=0\}
\end{align*}

Since elasticity is an equilibrium problem the Euler-Lagrange equation $\delta_{\boldsymbol{u}}\Psi = \boldsymbol{0}$ can be invoked. It also is the weak form of the governing equation for  elasticity:

\begin{align}
    \int_\Omega g(d) \frac{\partial\psi_e(\boldsymbol{C},d)}{\partial\boldsymbol{F}}:\nabla \boldsymbol{w}_{\boldsymbol{u}} \mathrm{d}V &= 0
    \label{eq:weakform_elast}
\end{align}

This implies the quasi-static approach to fracture, with elastic equilibrium imposed at each step under control of a loading parameter.  

However, the damage problem  is not at equilibrium, but evolves as a gradient flow shown in weak form \ref{eq:gradient_flow}, with the variational derivative in \ref{eq:euler_lagrange-damage} defining the right hand-side. A Macaulay bracket is introduced  to ensure the irreversibly of the damage.

\begin{equation}
   \int_\Omega \eta\dv{d}{t} w_d \mathrm{d}V = \left\langle - \frac{\delta \Psi(\boldsymbol{u},d)}{\delta  d} \right\rangle
    \label{eq:gradient_flow}
\end{equation}

The  viscosity parameter $\eta$ controls the kinetics of damage evolution. As mentioned in the discussion of the free energy density, \ref{eq:energy_density}, the terms that will be substituted into the Macaulay bracket are a balance between the bulk elastic free energy in the material and the damage energy, which includes the smooth fracture energy and the surface energy of the crack. As damage develops, the damage energy term  remains positive since the gradient term will increase as the smooth fracture energy remains smaller. This leaves the contributions of the bulk elastic free energy in the material, which could decrease if the material is unloaded since the degradation function cannot increase (implying damage reversal) to compensate for the reduction in the elastic energy. Therefore, a ``yield-like'' history term, $\mathcal{H}$, is introduced to drive damage without restricting the problem to the rate-dependent form. This function is defined using a crack driving state function $\widetilde{D}$ chosen to be the elastic energy $\widetilde{D} = \psi_e(\boldsymbol{C},d)$ \cite{miehe_phase_2010}:
 \begin{equation}
    \mathcal{H}(\boldsymbol{x}, t) = \text{max}_{s\in [0,t]} \widetilde{D}(\boldsymbol{x}, s | (\boldsymbol{C}, d)) \geq 0
    \label{eq:history_update}
\end{equation}

Using the history term, $\mathcal{H}$ instead of the undamaged elastic free energy density $\psi_e$ allows Macaulay bracket to be dispensed with from the gradient flow, \ref{eq:gradient_flow}. Using \ref{eq:euler_lagrange-damage} in \ref{eq:gradient_flow} with $\mathcal{H}$ substituted for $\psi_e$ leads to the rate-dependent, gradient flow weak form for damage evolution:
\begin{equation}
    \int_\Omega \eta \dv{d}{t} w_d \mathrm{d}V + \int_\Omega g'(d) \ \mathcal{H}(\boldsymbol{x},t) w_d \mathrm{d}V + \int_\Omega g_c c_s \left( \beta f'(d) w_d + \nabla w_d\cdot \nabla d \right) \mathrm{d}V = 0
    \label{eq:weakformdamage}
\end{equation}

In the strong form:
\begin{equation}
    \eta\dv{d}{t} = -g'(d)\mathcal{H}(\boldsymbol{x},t) - g_c c_s (\beta f'(d) - \nabla^2 d)
\end{equation}

A non-comprehensive list of crack-driving state functions appears in \ref{tab:history_functions}. The state function $\widetilde{D}$ was defined as the tensile component of the elastic energy, (D1), in the bulk.\cite{miehe_phase_2010} Alternate state functions were posed by considering the stress-strain relationship. Geelen et al. used a critical energy, (D2), defined by the  stress at which the material will begin to undergo damage.\cite{geelen_phasefield_2019} Along a similar approach, Raina and Miehe considered the introduction of damage based on the principal stresses, (D3), to model the isotropic failure surface in biological tissues.\cite{raina_phasefield_2016} Note that the critical energy, (D2), is a special condition of the energy function, (D1). 
\begin{table}[!ht]
\centering
\begin{tabular}{llll} \hline
Abbr. & Type                & $\widetilde{D}$ & Reference\\ \hline
(D1) & Energy              & $\widetilde{D}_1 = \psi_e$  & Miehe \cite{miehe_phase_2010} \\
(D2) & Critical Energy     & $\widetilde{D}_2 = \psi_{\text{crit}} + \left\langle\psi_e-\psi_{\text{crit}}\right\rangle$ & Geelen \cite{geelen_phasefield_2019} \\
(D3) & Principal Stress    & $\widetilde{D}_3 = \left\langle \sum_{i=1}^{3} \left(\frac{\langle \sigma_i \rangle }{\sigma_{\text{crit}}/\mathfrak{a}_i}\right)^2  -1 \right \rangle$  & Raina \cite{raina_phasefield_2016}   \\ 
\end{tabular}
\caption{Forms of the crack driving state function used as the history variable in the literature.}
\label{tab:history_functions}
\end{table}

\section{Finite element formulation}
\label{sec:fem}

The weak forms of elasticity and of damage evolution have been delineated in \ref{eq:weakform_elast} and \ref{eq:weakformdamage}.  In this section we outline the finite element formulation for solution of these equations in weak form.

\subsection{The Galerkin weak forms}

The finite dimensional weak form is written by replacing trial solutions and weighting functions with their finite-dimensional approximations $\boldsymbol{u}^h \in \mathcal{S}_u^h,\;\boldsymbol{w}_u^h \in \mathcal{V}_u^h, \; d^h\in \mathcal{S}_d^h, \; w_d^h\in \mathcal{V}_d^h$ in \ref{eq:weakform_elast} and \ref{eq:weakformdamage}, where functions in the spaces $\mathcal{S}_u^h$ and $\mathcal{S}_d^h$ satisfy the corresponding Dirichlet boundary conditions, while those in $\mathcal{V}_u^h$ and $\mathcal{V}_d^h$ satisfy zero Dirichlet boundary conditions. Using basis functions $N^A(\boldsymbol{X})$ for both the elasticity and the damage problems, these fields are:
\begin{alignat*}{3}
    \boldsymbol{u}^h(\boldsymbol{X})\bigg\vert_{\Omega_e} &&= \sum_{A=1}^{n_{ne}}N^A(\boldsymbol{X})\boldsymbol{b}_A, \quad\boldsymbol{w}^h_{{u}}(\boldsymbol{X})\bigg\vert_{\Omega_e} &&= \sum_{A=1}^{n_{ne}}N^A(\boldsymbol{X})\boldsymbol{c}_A\\
     d^h(\boldsymbol{X})\bigg\vert_{\Omega_e} &&= \sum_{A=1}^{n_{ne}}N^A(\boldsymbol{X})e_A, \quad w^h_d(\boldsymbol{X})\bigg\vert_{\Omega_e} &&= \sum_{A=1}^{n_{ne}}N^A(\boldsymbol{X})h_A
\end{alignat*}  
where the index $A\in\{1,\dots,n_{ne}\}$ ranges over the nodes of element $\Omega_e$, assumed to be the same for expansion of the paired fields $(\boldsymbol{u}^h,\boldsymbol{w}_u^h)$ and $(d^h,w^h_u)$. The nodal coefficients $\boldsymbol{b}_A,\boldsymbol{c}_A$ are  vectors  while $e_A,h_A$ are  scalars. This leads to the Galerkin weak form for the elasticity problem:

\begin{equation}
    \sum_{e=1}^{n_{el}}\int_{\Omega_e} \left( g(d)\boldsymbol{P}^h_{0_e}\left(\boldsymbol{1}+\nabla\sum\limits_{B=1}^{n_{ne}}N^B \boldsymbol{b_B}\right):\sum_{A=1}^{n_{ne}} \nabla N^A \boldsymbol{c}_A \right)\mathrm{d}V = 0
    \label{eq:disp_residual_galerkin}
\end{equation}
and for the damage problem:
\begin{align}
    &\sum_{e=1}^{n_{el}}\int_{\Omega_e}  \left(\eta \sum\limits_{B=1}^{n_{ne}}N^B\dot{e}_B + g^\prime\left(\sum\limits_{B=1}^{n_{ne}}N^B e_B\right)\mathcal{H}(\boldsymbol{u}^h) + \beta g_c c_s f^\prime\left(\sum\limits_{B=1}^{n_{ne}}N^B e_B\right)\right) \sum_{A=1}^{n_{ne}}N^A h_A \mathrm{d}V\nonumber\\
    &\phantom{\sum_{e=1}^{n_{el}}}+\int_{\Omega_e}\left( g_c c_s \nabla \left(\sum\limits_{B=1}^{n_{ne}}N^B e_B \right)\cdot \nabla \left(\sum_{A=1}^{n_{ne}}N^A h_A\right) \right) \mathrm{d}V = 0
    \label{eq:damage_residual_galerkin} 
\end{align}
where the integrals in \ref{eq:weakform_elast} and \ref{eq:weakformdamage} have been replaced with summation over the $n_{el}$ element integrals. 

Introducing the global nodal vectors for displacement $\boldsymbol{b}$, and damage $\boldsymbol{e}$, and invoking the fact  the finite-dimensional weak forms hold for all $\boldsymbol{w}^h_u \in \mathcal{V}_u^h$ and $w_d^h \in \mathcal{V}_d^h$ the global residual vectors, $\boldsymbol{R}_u$ and $\boldsymbol{R}_d$, are obtained by the finite element  assembly operator over the  elements, $\femsum_{e=1}^{n_{el}}{}$.  
\begin{equation}
    \boldsymbol{R}_{\boldsymbol{u}}(\boldsymbol{b},\boldsymbol{e}) = \femsum_{e=1}^{n_\text{el}}{\int_{\Omega_e} \left( g(d) \langle \boldsymbol{P}_{0_e}\nabla N^1 ,\dots,\boldsymbol{P}^h_{0_e}\nabla N^{n_{\text{ne}}}\rangle^\text{T} \right)\mathrm{d}V}
    \label{eq:disp_residual_vector}
\end{equation}
and introducing the vector of basis functions on the element $e$: $\boldsymbol{N}^e=\langle N^1,\dots,N^{n_{ne}}\rangle^T\in\mathbb{R}^{n_\text{ne}}$,

\begin{align}
    \boldsymbol{R}_d(\boldsymbol{b},\boldsymbol{e}) =&\femsum_{e=1}^{n_\text{el}}{\int_{\Omega_e}  \boldsymbol{N}^e\left(\eta \sum\limits_{B=1}^{n_{ne}}N^B\dot{e}_B + g^\prime\left(\sum\limits_{B=1}^{n_{ne}}N^B e_B\right)\mathcal{H}(\boldsymbol{b},\boldsymbol{e}) + \beta g_c c_s f^\prime\left(\sum\limits_{B=1}^{n_{ne}}N^B e_B\right)\right)  \mathrm{d}V}\nonumber\\
    &+\femsum_{e=1}^{n_\text{el}}{\int_{\Omega_e}\nabla\boldsymbol{N}^e\left( g_c c_s \nabla \left(\sum\limits_{B=1}^{n_{ne}}N^B e_B \right) \right) \mathrm{d}V}
    \label{eq:damage_residual_vector} 
\end{align}

where, with an abuse of notation in favor of brevity, in the finite dimensional weak form the history term is parameterized as:
\begin{equation}
\begin{split}
    \mathcal{H}(\boldsymbol{b}(t),\boldsymbol{e}(t)) &= \max_{(\boldsymbol{b}(s),\boldsymbol{e}(s)),s\in[0,t]}\tilde{D}(\boldsymbol{b}(s),\boldsymbol{e}(s))
    \label{eq:H}
\end{split}
\end{equation}

\section{Solution scheme}
\label{sec:fwdsoln}
The phase-field fracture models that appeared early in the literature were solved using monolithic schemes as in Bourdin.\cite{bourdin_numerical_2000} In this approach, the full regularized functional, with both displacement and damage, is minimized using an iterative scheme of successive minimizations for the damage and then the displacement. Convergence may not be guaranteed with this scheme or can be slow since the energy functional is not convex in general. Introduction of the history variable by Miehe et al. enabled the algorithmic decoupling of the governing equations.\cite{miehe_phase_2010} This staggered approach solves a system of equations in which one field, e.g. $d^h$ is held fixed when solving for the other field, $\boldsymbol{u}^h$, by enforcing the elasticity residual $\boldsymbol{R}_u = \boldsymbol{0}$ in \ref{eq:disp_residual_vector} via a nonlinear solution scheme. Next, the residual for damage evolution is enforced: $\boldsymbol{R}_d = \boldsymbol{0}$ in \ref{eq:damage_residual_vector}, while holding $\boldsymbol{u}^h$ fixed to solve for $d^h$,  This operator split scheme performs well, being more robust compared to a monolithic scheme.

The algorithm used in this work is shown in Algorithm \ref{alg:staggered_scheme}. First, the state and history variables are initialized to zero. In the time loop, the monotonically increasing displacement-driven loading is evaluated. Then, a condition is held until the iterations solving the displacement and damage residuals meet a certain error tolerance. The displacement residual is solved first, with the introduction of a small epsilon value, $\epsilon$, in the damage field to ensure numerical stability by preventing zero stiffness. An additional constraint is also added to the residual that the damage does not evolve $d-d_{temp}$. Temporary variables are updates for the state and history functions and then the damage residual is solved along with a similar constraint that the displacement does not evolve, $u=u_{temp}$. Finally, the error for the sub-iteration is evaluated, verifying that the temporary displacement and damage fields match the fields evaluated within the loop. If the tolerance is met, the loop will exit and move to the next iteration.

\begin{algorithm}
\caption{Implicit staggered solution scheme for phase-field fracture from time $t_n$ to $t_{n+1}$}\label{alg:staggered-scheme}
\begin{algorithmic}[1]
\State $\vec{u}=\vec{0}, d=0$
\State $\vec{u}_n=\vec{u}$, $d_n=d$
\For{t=[0,N]}
    \State $\overline{u} = disp_{rate}*t$
    \State $\vec{u}_{temp}=\vec{u}_n, d_{temp}=d_n$
    \While{$err_{iter} > tol$ and \texttt{iter} > \texttt{max\_iter}}
        \State Solve $R_u = 0$ while $d=d_{temp}$
        \State $\vec{u}_{temp} = \vec{u}$
        \State Solve $R_d = 0$ while $\vec{u}=\vec{u}_{temp}$, $\mathcal{H}$ from (\ref{eq:H}) 
        \State $d_{temp}=d$
        \State $err_{iter} = \sqrt{\norm{\vec{u}-\vec{u}_{temp}}_2 + \norm{d-d_{temp}}_2}$ 
    \EndWhile  \label{subiter_loop}
    \State $\vec{u}_n = \vec{u}$, $d_n=d$
\EndFor
\end{algorithmic}
\label{alg:staggered_scheme}
\end{algorithm}
\section{System identification}
\label{sec:inference}

The phase-field approach to modeling fracture has been developed with variations to accommodate different materials and damage modes. While these models have been compared to experimental results independently and compared to each other in review literature, the process of selecting the optimal model for a given material and loading condition has not been explored, and is the focus of this communication. We begin with the ansatz for energy density, elastic energy, and the assumption that all elastic properties $(\lambda,\mu,\theta_1, \dots, )$ of the material are known. Our aim is to infer the phase field model for fracture by identifying the damage parameters along with the degradation function, smooth fracture energy function, and the history variable update approach. 

\begin{figure}[H]
    \centering
    \includegraphics[width=\linewidth]{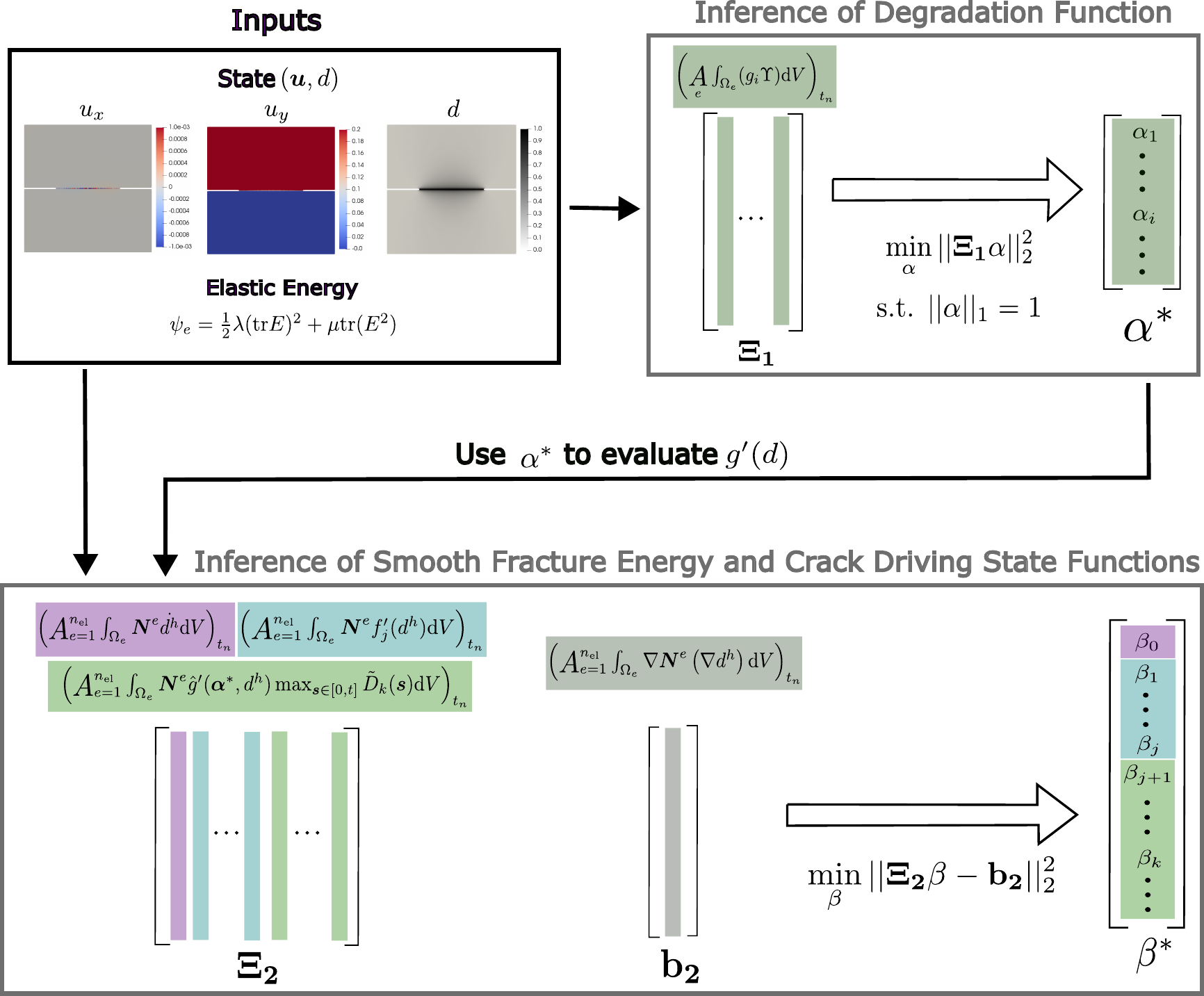}
    \caption{Schematic of the variational system identification algorithm for phase field fracture. The inputs are the candidate material models, and the data are the displacements $(u_x,u_y)$ as well as the scalar phase field $d$. These are used to evaluate the matrices of candidate operators $\Xi_1$ and $\Xi_2$. The degradation function is identified first, so that the resulting $\boldsymbol{\alpha^{\ast}}$ is used in evaluating $\boldsymbol{\beta}^{\ast}$.}
    \label{fig:vsi_schematic}
\end{figure}

In order to set up the inference problem in the variational form, we need to introduce  candidate operators within the weak form. System identification then proceeds in two stages: first, inferring the degradation function using \ref{eq:disp_residual_vector}, and second, inferring the smooth fracture energy function and history update using \ref{eq:damage_residual_galerkin}. In both inference steps, the data is loaded, and the boundary conditions are applied. Then, the individual terms of the first and second residual vectors, $\boldsymbol{R}_u$ and $\boldsymbol{R}_d$, respectively, are evaluated on the interior nodes and saved into the matrices/vector: $\boldsymbol{\Xi}_1$ detailed in \cref{sec:first_inference}, and $\boldsymbol{\Xi}_2$ and $\boldsymbol{b}_2$ detailed in \cref{sec:second_inference}. Following inference of the degradation function, the resulting coefficients are represented in the matrix $\boldsymbol{\alpha}^\ast$. These inferred coefficients are then used in the second inference problem to identify the smooth fracture energy and history update, represented in the matrix $\boldsymbol{\beta}^\ast$. A schematic representation of this approach appears in \ref{fig:vsi_schematic}.


\subsection{Inference of the degradation function} 
\label{sec:first_inference}

Given data for the field $d$, the true composite degradation function $g(d)$ is written in parametric form as $\hat{g}(\boldsymbol{\alpha};d)$, \ref{eq:parametric_degradation}, with unknown coefficient vector $\boldsymbol{\alpha}$ containing  components $\alpha_i$ and corresponding candidate degradation functions $g_i(d)$, selected from \ref{tab:degradation_functions}.

\begin{equation}
        \hat{g}(\boldsymbol{\alpha};d) = \sum_i\alpha_ig_i(d) 
    \label{eq:parametric_degradation}
\end{equation}

The $g_i(d)$ are assumed to be linearly independent. Substituting the  parameterized degradation function $\hat{g}(\boldsymbol{\alpha};d)$ in the residual vector for the elasticity problem, \ref{eq:disp_residual_vector}, we have:
\begin{equation}
    \boldsymbol{R}_u(\boldsymbol{b},\boldsymbol{e}) = \femsum\limits_e{\int_{\Omega_e} \left( \sum_i\alpha_ig_i(d)  \langle\boldsymbol{P}_{0_e}\nabla N^1,\dots \boldsymbol{P}_{0_e}\nabla N^{n_\text{ne}}\rangle^\text{T} \right)\mathrm{d}V} = 0
    \label{eq:displacement_residual_vector_parameterized}
\end{equation}

By using data for the fields $\boldsymbol{u}^h_n,d^h_n$ at times $\{t_0,\dots,t_n,\dots\}$ converted to nodal vectors $\boldsymbol{b}_n,\boldsymbol{e}_n$ following \cref{sec:fem}, \ref{eq:displacement_residual_vector_parameterized} can be expanded into block rows corresponding to each time instant. For brevity, $\langle\boldsymbol{P}_{0_e}\nabla N^1,\dots \boldsymbol{P}_{0_e}\nabla N^{n_\text{ne}}\rangle^\text{T}$ is represented with $\Upsilon$ in \ref{eq:linear_vsi_first_matrices}.

\begin{equation}
    \boldsymbol{\Xi_1} =  \begin{bmatrix} 
        \left(\femsum\limits_e{\int_{\Omega_e} (g_1\Upsilon)\mathrm{d}V}\right)_{t_0} && \left(\femsum\limits_e{\int_{\Omega_e} (g_2\Upsilon)\mathrm{d}V}\right)_{t_0} && \dots \\  
        \vdots && \vdots  && \vdots \\
        \left(\femsum\limits_e{\int_{\Omega_e} (g_1\Upsilon)\mathrm{d}V}\right)_{t_n} && \left(\femsum\limits_e{\int_{\Omega_e} (g_2\Upsilon)\mathrm{d}V}\right)_{t_n} && \dots \\ 
        \vdots && \vdots && \vdots 
    \end{bmatrix}, 
    \hspace{3em}
    \boldsymbol{\alpha} = \begin{bmatrix} \alpha_1 \\ \alpha_2 \\ \vdots \end{bmatrix}
    \label{eq:linear_vsi_first_matrices}
\end{equation}

The composite degradation function can now be obtained by solving an optimization problem over the $\alpha_i$ coefficients, as  in \ref{eq:linear_vsi_first_min}. The properties of the degradation function discussed in the context of \ref{eq:degradation_function_properties} must also hold for the sum of candidate functions $g_i(d)$ in the parameterized form. Therefore, we have the properties $\hat{g}(\boldsymbol{\alpha};0)=1$ and $\hat{g}(\boldsymbol{\alpha};1)=0$, subject to the constraint $\sum_i\alpha_i = 1$. 

\begin{equation}
    \begin{split}
    \{\alpha_1^{\ast}, \cdots, \alpha_n^{\ast}\} = \argmin_{\boldsymbol{\alpha}} ||\boldsymbol{\Xi_1\alpha}||_2^2 \hspace{1em} &\textrm{s.t.}
    \hspace{1em} \sum_i \alpha_i = 1
    \end{split}
    \label{eq:linear_vsi_first_min}
\end{equation}

Standard solution approaches such using as the pseudo-inverse and ridge regression cannot be exploited due to the constraint. Therefore, the Python package \texttt{CVXPY} was used to solve this problem with computational efficiency. 


\subsection{Inference of smooth fracture energy and crack driving state functions}
\label{sec:second_inference}

In a similar approach to the  inference of the composite degradation function outlined in \cref{sec:first_inference}, the smooth fracture energy function and history update approach $\mathcal{H}(\boldsymbol{x},t)$ will be considered in parametric form. Given data for the field $d$, the composite smooth fracture energy $\hat{f}(\boldsymbol{\beta};d)$ in \ref{eq:parametric_geometric} is a function of unknown coefficients $\beta_j$ and the candidate functions $f_j(d)$, selected from those used in the literature shown in \ref{tab:geometric_functions}. The index $j$ begins at 1 to account for the existing constant $\beta_0$ and goes up to the total number of smooth fracture energy functions $n_{sfe}$. 

\begin{equation}
    \hat{f}(\boldsymbol{\beta};d) = \sum_{j=1}^{n_{sfe}} \beta_j f_j(d)
    \label{eq:parametric_geometric}
\end{equation}

For the history update approach, the parametric form is specifically considered for the crack driving state function $\widetilde{D}$ to remain numerically stable for history parameterization $\mathcal{H}$ with the Macaulay bracket. Therefore, $\hat{D}(\boldsymbol{\beta};\boldsymbol{s})$, shown in \ref{eq:parametric_history} is a function of unknown coefficients $\beta_k$ and the state functions $\tilde{D}_k(\boldsymbol{s})$ listed in \ref{tab:history_functions}. The state $\boldsymbol{s}$ is a concatenation of the nodal degrees of freedom $(\boldsymbol{b},e)$. The index $k$ begins after the number of smooth fracture energy functions, $k_0=n_{sfe}+1$, and ends at the total number of crack driving state functions $k_0 + n_{cdf}$.

\begin{equation}
    \hat{D}(\boldsymbol{\beta};\vec{s}) = \sum_{k=k_0}^{k_0+n_{cdf}} \beta_k \tilde{D_k}(\boldsymbol{s})
    \label{eq:parametric_history}
\end{equation}

The coefficients for these two parameterized functions can be identified using the damage residual from \ref{eq:damage_residual_galerkin}. To set up the inference problem, we divide through by the coefficients $g_c$ and $c_s$. We also substitute the smooth fracture energy $\hat{f}(\boldsymbol{\beta};d)$ and the crack driving state function, $\hat{D}(\boldsymbol{\beta};\boldsymbol{s})$. For readability, we rename the scaled coefficient as $\beta_0$ as shown in \ref{eq:damage_residual_vector_parameterized}.

\begin{equation}
    \begin{split}
    \widetilde{\boldsymbol{R}}_d(\boldsymbol{b},\boldsymbol{e}) = &\femsum_{e=1}^{n_\text{el}}{\int_{\Omega_e} \boldsymbol{N}^e\Bigg[\underbrace{\frac{\eta}{g_c c_s}}_{\beta_0} \sum\limits_{B=1}^{n_{ne}}N^B\dot{e}_B + \sum_j \beta_j f_j'\left(\sum\limits_{B=1}^{n_{ne}}N^B e_B\right)}
    \\ &\hspace{5em} + \hat{g}^\prime\left(\boldsymbol{\alpha}^\ast,\sum\limits_{B=1}^{n_{ne}}N^B e_B\right)\sum_k \underbrace{\frac{1}{g_c c_s}}_{\beta_k} \max_{\boldsymbol{s}\in[0,t]}\tilde{D}_k(\boldsymbol{s}) \Bigg]  \mathrm{d}V
    \\ &+ \femsum_{e=1}^{n_\text{el}}{\int_{\Omega_e}\nabla\boldsymbol{N}^e\left( \nabla \left(\sum\limits_{B=1}^{n_{ne}}N^B e_B \right) \right) \mathrm{d}V}
    \end{split}
    \label{eq:damage_residual_vector_parameterized}
\end{equation}

The coefficients in the final damage residual appear in \ref{eq:damage_residual_constants}. Recall that the values for $c_s$ is determined based on existing models for the regularized fracture energy summarized in \ref{tab:crack_energy_models}. The value for the viscosity $\eta$ is determined in order to produce the best forward simulation data and the constant $g_c$ is chosen based on the fracture toughness of the material. Note that the $\beta_k$ coefficients are combinations of $g_c$ and $c_s$, and they are the parameters that will be identified for the parameterized form of the history update approach.
\begin{equation}
    \beta_0 = \frac{\eta}{g_c c_s} \in \mathbb{R}^+\cup\{0\}, \hspace{2em} \beta_j\in \mathbb{R}^+\cup\{0\}, \hspace{2em} \beta_k = \frac{1}{g_c c_s}\in \mathbb{R}^+\cup\{0\}
    \label{eq:damage_residual_constants}
\end{equation}

For brevity, we re-write \ref{eq:damage_residual_vector_parameterized} using the notation for the discretized damage $d^h$ and derivative $\dot{d^h}$:
\begin{equation}
    \begin{split}
    \widetilde{\boldsymbol{R}}_d(\boldsymbol{b},\boldsymbol{e}) = &\femsum_{e=1}^{n_\text{el}}{\int_{\Omega_e} \boldsymbol{N}^e\Bigg[\beta_0 \dot{d^h} + \sum_j \beta_j f_j'(d^h) + \hat{g}^\prime(\boldsymbol{\alpha}^\ast,d^h)\sum_k\beta_k \max_{\boldsymbol{s}\in[0,t]}\tilde{D}_k(\boldsymbol{s})} \Bigg]  \mathrm{d}V
    \\ &+ \femsum_{e=1}^{n_\text{el}}{\int_{\Omega_e}\nabla\boldsymbol{N}^e\left( \nabla d^h \right) \mathrm{d}V}
    \end{split}
    \label{eq:damage_residual_vector_parameterized_short}
\end{equation}

Expanding this form and factoring out the constants, we can write:
\begin{equation}
    \begin{split}
    \widetilde{\boldsymbol{R}}_d(\boldsymbol{b},\boldsymbol{e}) = &\beta_0\underbrace{\femsum_{e=1}^{n_\text{el}}\int_{\Omega_e} \boldsymbol{N}^e \dot{d^h} \mathrm{d}V}_{\boldsymbol{K}^1} + \sum_j \beta_j\underbrace{\femsum_{e=1}^{n_\text{el}}\int_{\Omega_e} \boldsymbol{N}^e  f_j'(d^h) \mathrm{d}V}_{\boldsymbol{K}_j^2} \\
    &+ \sum_k\beta_k\underbrace{\femsum_{e=1}^{n_\text{el}}\int_{\Omega_e} \boldsymbol{N}^e \hat{g}^\prime(\boldsymbol{\alpha}^\ast,d^h) \max_{\boldsymbol{s}\in[0,t]}\tilde{D}_k(\boldsymbol{s}) \mathrm{d}V}_{\boldsymbol{K}_k^3} + \underbrace{\femsum_{e=1}^{n_\text{el}}{\int_{\Omega_e}\nabla\boldsymbol{N}^e\left( \nabla d^h \right) \mathrm{d}V}}_{\boldsymbol{K}^c}
    \end{split}
\end{equation}

Since $f_j(d^h)$ and $\tilde{D}_k(\boldsymbol{s})$ are candidate functions for $\hat{f}$ and $\hat{D}$, respectively, \ref{eq:damage_residual_vector_parameterized} may be written as a linear system of equations with the matrices shown in \ref{eq:linear_vsi_second_matrices}. As for the inference of degradation function, by using data for the fields $\boldsymbol{u}^h_n,d^h_n$ at times $\{t_0,\dots,t_n,\dots\}$ converted to nodal vectors $\boldsymbol{b}_n,\boldsymbol{e}_n$ following \cref{sec:fem}, \ref{eq:damage_residual_vector_parameterized} can be cast into block rows corresponding to each time instant. The term on the second line of \ref{eq:damage_residual_vector_parameterized} is not a function of unknown parameters, so it may be used to define the label $\boldsymbol{b}_2$. 

\begin{equation}
    \boldsymbol{\Xi_2} = \begin{bmatrix}
        \left(\boldsymbol{K}^1\right)_{t_0} & \dots & \left(\boldsymbol{K}_j^2\right)_{t_0} & \dots & \left(\boldsymbol{K}_k^3\right)_{t_0} & \dots \\ 
        \vdots & \vdots & \vdots & \vdots & \vdots & \vdots \\
        \left(\boldsymbol{K}^1\right)_{t_n} & \dots & \left(\boldsymbol{K}_j^2\right)_{t_n} & \dots & \left(\boldsymbol{K}_k^3\right)_{t_n} & \dots \\
        \vdots & \vdots & \vdots & \vdots & \vdots & \vdots
    \end{bmatrix}
    \hspace{3em} \boldsymbol{\beta} = \begin{bmatrix} \beta_0 \\ \vdots \\ \beta_j \\ \vdots \\ \beta_k \\ \vdots \end{bmatrix}
     \hspace{2em} \boldsymbol{b_2} = \begin{bmatrix} \left(\boldsymbol{K}^c \right)_{t_0} \\ \vdots \\\left(\boldsymbol{K}^c \right)_{t_n} \\ \vdots \end{bmatrix}
\label{eq:linear_vsi_second_matrices}
\end{equation}

This results in the standard least squares minimization:
\begin{equation}
    \{\boldsymbol{\beta}^\ast\} = \argmin_{\boldsymbol{\beta}}\vert\vert \boldsymbol{\Xi_2\beta} - \boldsymbol{b_2} \vert\vert_2^2 
    \label{eq:linear_vsi_second_min}
\end{equation}

This standard least squares minimization could be solved using the pseudo-inverse ($\boldsymbol{\beta}^\ast = (\boldsymbol{\Xi}_2^T\boldsymbol{\Xi}_2)^{-1}\boldsymbol{\Xi}_2^T\boldsymbol{b}_2$). However, \texttt{CVXPY} was used for computational efficiency and to keep the implementation consistent with that for the inference of the degradation function using \ref{eq:linear_vsi_first_min}.


\section{Synthetic data generation}
\label{sec:datagen}
In this first communication on the topic, our goal is to demonstrate the viability of the inference framework using synthetic data. The scope of our initial study is restricted to a single boundary value problem. However, given this work's emphasis on data-driven modeling, we have investigated the robustness of the approach for data subsampling and in the presence of noise. Th data were generated via forward simulations of phase field fracture using the \texttt{FEniCS} library. 

\subsection{The forward and inference problem: Plain strain fracture in a double edge notched plate}

\begin{figure}[!ht]
    \centering
    \includegraphics[width=0.8\linewidth]{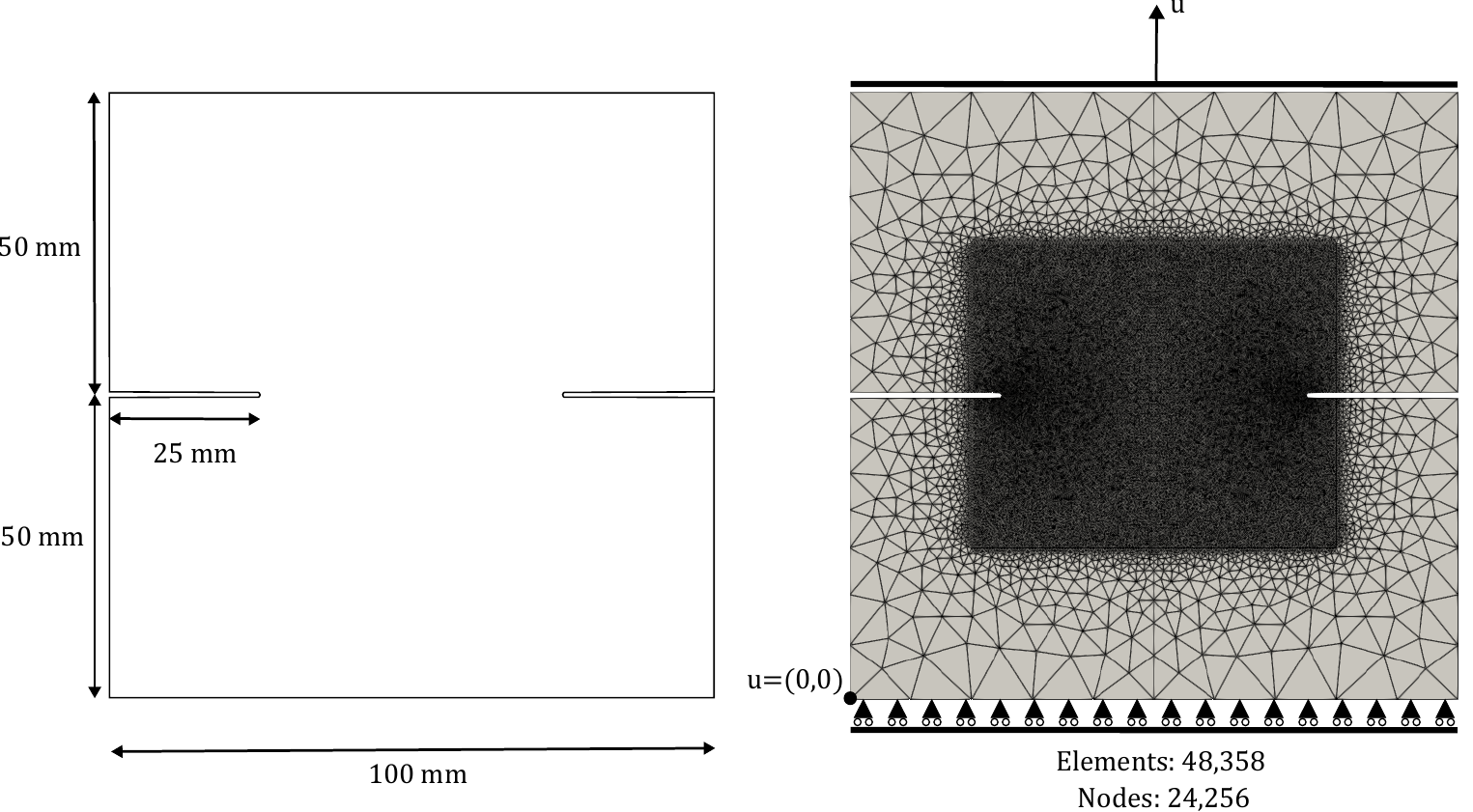}
    \caption{Schematic of the double edge notched geometry (left) and the finite element mesh with loading conditions (right) used in this work.}
    \label{fig:example1_schematic}
\end{figure}

For the benchmark boundary value problem we used a thick square plate with double edge notches, following Suh and Sun.\cite{suh_opensource_2019} See \ref{fig:example1_schematic}. Uniaxial loading was applied with a vertical displacement prescribed along the top boundary at a rate of $1$ mm/s and zero vertical displacement on the bottom boundary. The bottom left corner was fixed, $\boldsymbol{u}= \boldsymbol{0}$, to prevent rigid body motion. As mentioned in \cref{sec:elastic_energy}, the Saint-Venant Kirchoff model of hyperelasticity was adopted with Lam\'{e} parameters $\lambda = 8.33$ GPa, $\mu = 12.5$ GPa (Young's modulus $E=30$ GPa, Poisson ration $\nu=0.2$). Plane strain conditions were assumed. The fracture parameters were chosen to be $g_c=0.1$ N/mm, $\ell_c=0.75$ mm, and $\psi_{\textrm{crit}}=1.0$ kJ/m\textsuperscript{3}. Model P5 from \ref{tab:crack_energy_models} was used for the $c_s$ and $\beta$ coefficients. A viscosity of $\eta=5\times 10^{-3}$ GPa-sec/mm was used in order to produce well-resolved solutions in time with a reasonable time step: $\Delta t = 1\times 10^{-4}$ s. 

\subsubsection{Functions for VSI}
\label{sec:functions_for_vsi}

A summary of the different functions considered in this preliminary work appears in \ref{tab:VSI_combinations}. The quadratic and quasi-quadratic forms of the degradation function, $g(d)$, were selected since they are popular in the literature and keep the smooth fracture function inference problem linear. While a wider array of degradation functions could well be considered in this work, the two functions are sufficient to test the inference framework. The three main variants of smooth crack surface energy functions, $f(d)$, were considered. The linear and single well forms are the most popular from the literature, however, the double well function provides an interesting challenge for the inference framework since the damaged and undamaged states may have similar energies in the two wells corresponding to the damaged and undamaged phases. Finally, only the total elastic energy was considered for the crack driving state function, $\tilde{D}$, from the candidate functions listed in \ref{tab:history_functions}. This is the most commonly employed criterion with the appeal of being a direct consequence of the variational, gradient flow approach to phase field fracture. The critical energy formulation is a subset of the total elastic energy and is therefore subsumed by the latter. The stress based criteria was not used considered because it requires additional numerical treatment.

\begin{table}[h!]
\centering
\begin{tabular}{l|l|l}
\multicolumn{1}{c}{Degradation ($\hat{g}$)}          & \multicolumn{1}{c}{Smooth Fracture Energy ($\hat{f}$)} & \multicolumn{1}{c}{History ($\hat{D}$)} \\ \hline
$g_1(d) = (1-d)^2$                              & $f_1(d) = d$          & $D_1 = \psi_e$ \\
$g_2(d) = \frac{(1-d)^2}{(1-d)^2 + m*d(1+p*d)}$ & $f_2(d) = d^2$        &  \\
& $f_3(d) = d^2(1-d)^2$ &   
\end{tabular}
\caption{Phase field fracture functions used in the inference study. In the quasi-quadratic degradation function, $g_2(d)$, $m=50$ and $p=10$ are fixed parameters selected based on Suh and Sun.\cite{suh_opensource_2019}}
\label{tab:VSI_combinations}
\end{table}

\subsection{Forward data}

\subsubsection{Regions of the stress-strain curve}
\label{sec:regions_of_stressStrain_curve}

\begin{figure}[h!]
    \centering
    \includegraphics[width=\linewidth]{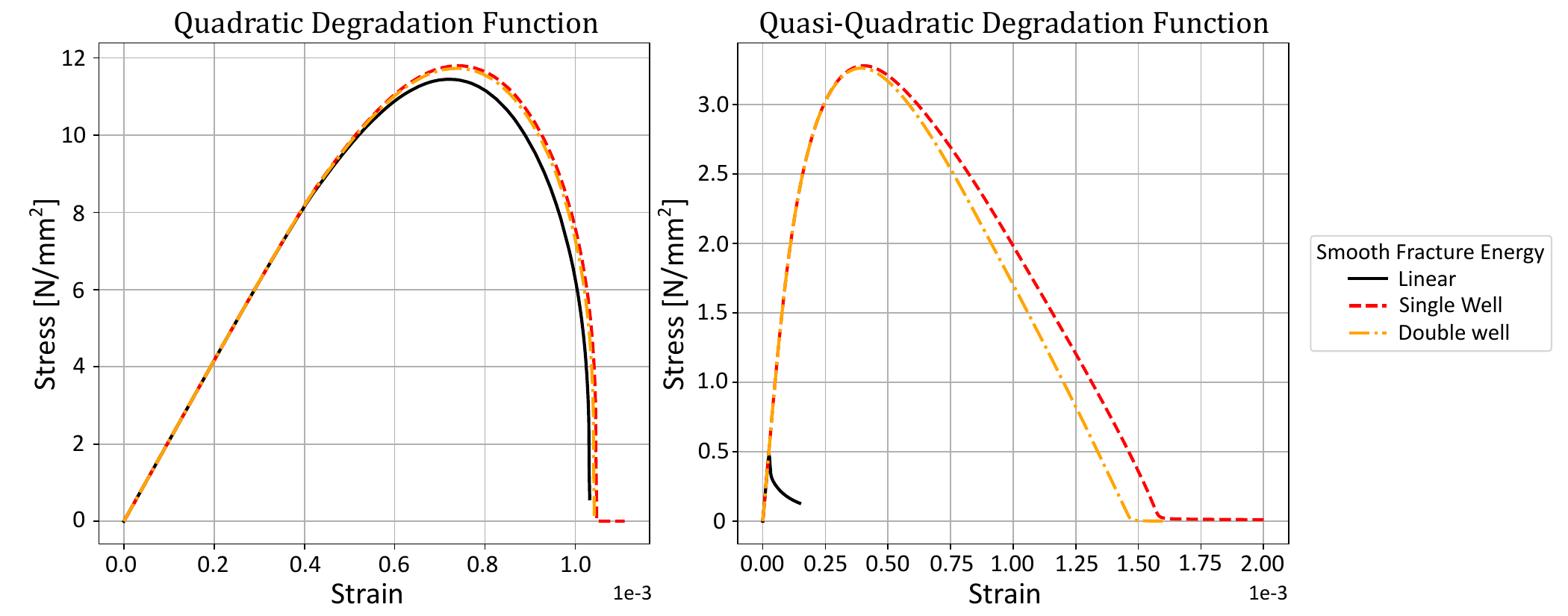}
    \caption{Stress-strain curves for quadratic (left) and quasi-quadratic (right) degradation functions with different smooth fracture energy functions.}
    \label{fig:quad-and-quasiQuad_loadstrain_comparison}
\end{figure}

A general overview of the six combinations of functions is shown in \ref{fig:quad-and-quasiQuad_loadstrain_comparison} through a comparison of the stress-strain curve. This curve is generated by evaluating the nominal stress over the top boundary, and the nominal stain.

For brevity, we refer to the different cases by the combinations of degradation and smooth fracture energy function; e.g., ``linear, quadratic''. It is important to note that the linear, quasi-quadratic simulation has a smaller peak stress and slower damage propagation, thus significantly impacting the shape of the stress-strain curve. This combination of functions also required a much smaller time step, increasing computational cost. The limitations of this linear smooth fracture energy function on the covexivity of the problem have been discussed in Svolos et al.\cite{svolos_convexity_2023}

The degradation function, $g(d)$, has a large effect on the peak nominal stress that is reached as well as the overall shape of the curve. Despite the differences in shape, the stress-strain curve can almost always be split into 4 main regions: ``elastic'' deformation of nearly linear response, damage initiation, damage propagation, and failure. An example of the cutoffs for each of these regions is shown in \ref{fig:load_strain_regions} and the state of damage at each cutoff point is shown in \ref{fig:quadratic_damage_comparison} using the quadratic degradation functions. The first point, A, is not shown since this is the starting point of the simulation. The last, E, is also not included since it is after the point at which the material has completely failed, and the data at this stage is not available for all simulations.

\begin{minipage}{\textwidth}
    \begin{minipage}[m]{0.49\textwidth}
        \centering
        \includegraphics[width=\linewidth]{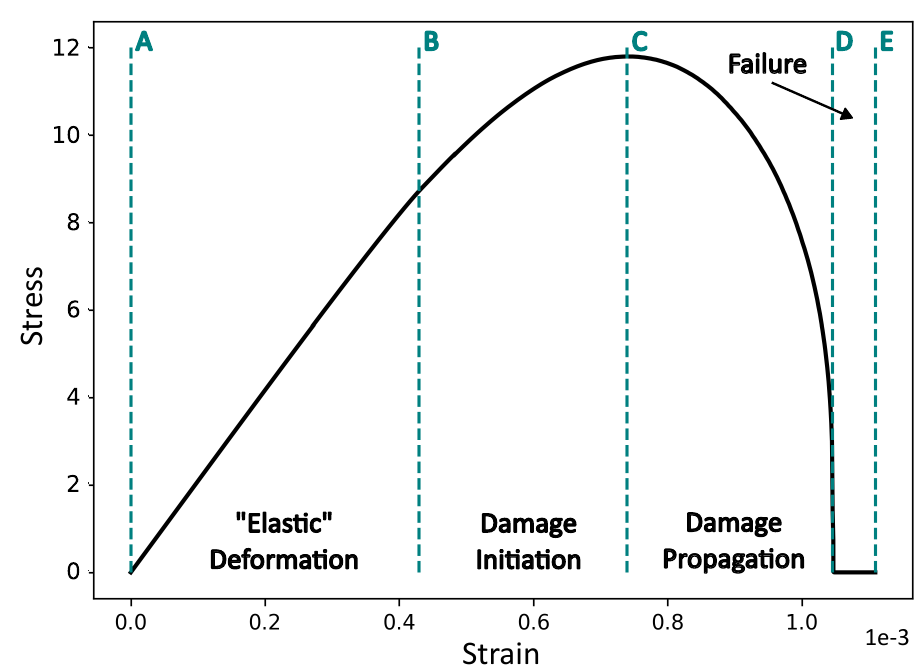}
        \captionsetup{width=0.85\textwidth}
        \captionof{figure}{A representative example (quadratic, single well) of the cutoff points selected for regions of the stress-strain curve.}
        \label{fig:load_strain_regions}
    \end{minipage}
    \hfill
    \begin{minipage}[m]{0.49\textwidth}
        \centering
        \includegraphics[width=\linewidth]{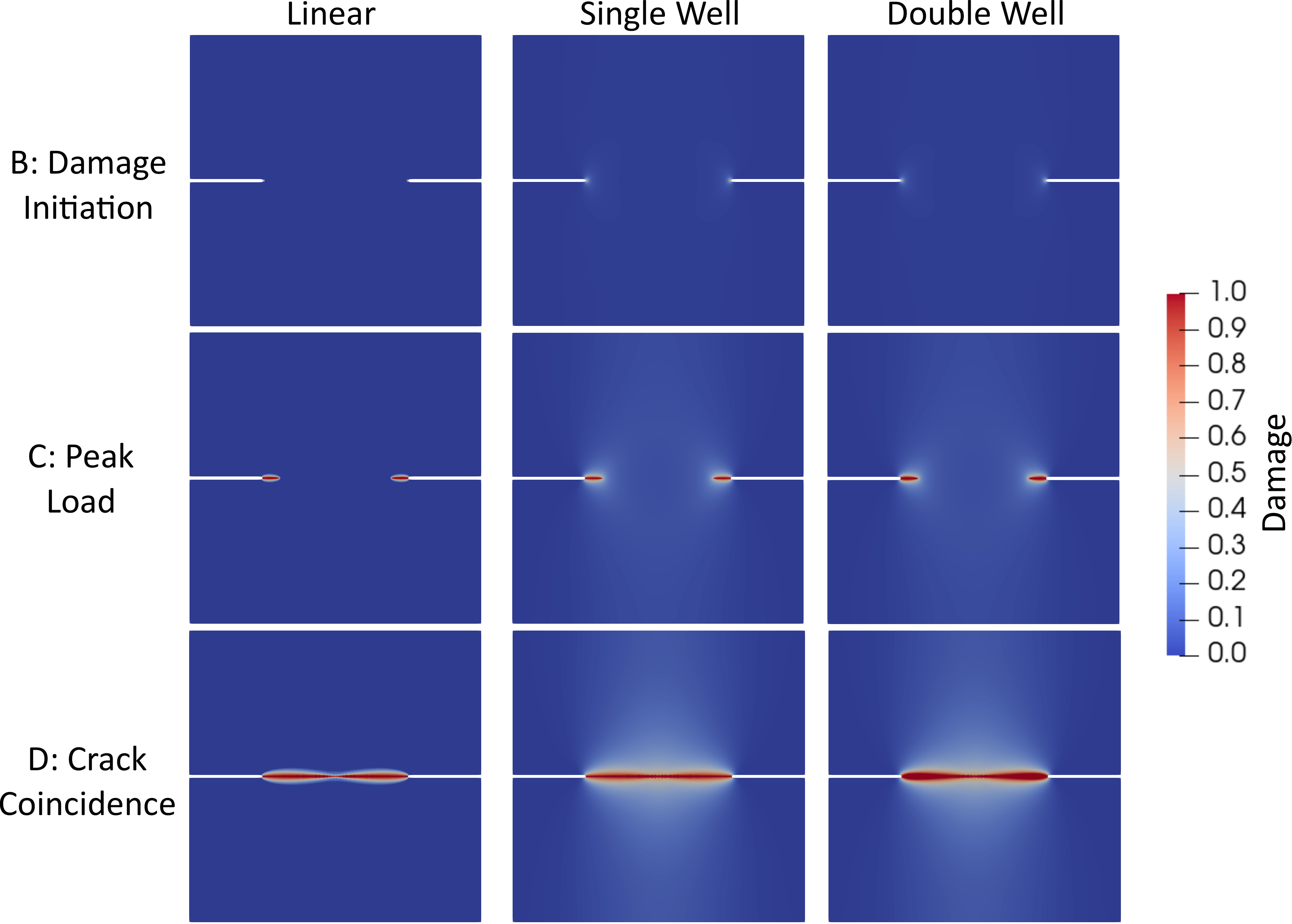}
        \captionsetup{width=0.8\linewidth}
        \captionof{figure}{Damage at region cutoffs denoted in \ref{fig:load_strain_regions}, using quadratic degradation function and comparing smooth fracture energy functions: linear, single well, and double well.}
        \label{fig:quadratic_damage_comparison}
    \end{minipage}
\end{minipage}

These regions delineate the physics of the fracture initiation and propagation. In the ``elastic'' region, $d \ll 1$. This can be seen in the first row for point B in \ref{fig:quadratic_damage_comparison} which shows the end of the ``elastic'' regime. With no crack surface at this stage, the smooth fracture energy and crack driving state function cannot be identified. Despite this limitation, the low level of damage should still be sufficient for identifying the degradation function using the elasticity residual. However, this approach will only work with damage (phase) field data from simulations. Damage in the form of micro-cracking or micro-voiding may not be prominent in experimental images over this regime. 

Once the simulation progresses beyond point B, damage initiates, increasing until the values reach 1 at which point a crack has formed (B to C in \ref{fig:quadratic_damage_comparison}). At this stage, point C, the crack surface appears and the diffuse field of damage $d > 0$ is visible. There is sufficient data to successfully infer both the degradation function and the smooth fracture energy. There is a small amount of damage propagation in this region while the crack surface forms and the diffuse effects of the crack begin to appear in the bulk of the material. This is clear in \ref{fig:quadratic_damage_comparison}, where the crack surface appears similar between smooth crack energy functions. However, the damage in the bulk is different. While the single and double well fields look similar in this view, the availability of high dimensional full-field data help identify the subtle differences between these damage fields.

In the damage propagation regime, C to D, the crack advances through the material, significantly affecting the elastic energy and reducing the load capacity. Similar to the initiation region, there is information sufficient to infer the degradation and smooth fracture energy functions. This is also the experimental regime where the crack is prominent and propagating. At the end of this regime, \ref{fig:quadratic_damage_comparison} point D, the differences between the single and double well smooth fracture energy functions are also more evident, suggesting that inference of the smooth fracture energy and crack driving state functions may be more precise, even with lower resolution in the spatio-temporal data.

In the failure regime, D to E, the crack has advanced through the sample and the forward simulations only continue because the phase field model of damage allows setting a tolerance $g(d) \ge \varepsilon > 0$ for the degradation function. In practice, however, in the limit $g(d) \to \varepsilon$  smaller time steps are required to advance the computation due to the near-elastic instability, and convergence is not always achieved. In an experiment, this is the fully failed regime and no data would be available.

\subsection{Inference coefficient summary}
\label{sec:inference_coefficient_summary}

The selected functions in \cref{sec:functions_for_vsi} along with the redefinition of coefficients in \ref{eq:damage_residual_constants} leads to the list in \ref{tab:coefficient_summary}. These form the ground truth model with which forward computations are run to generate data. For the degradation function, $\alpha_0$ and $\alpha_1$ will be equal to 0 or 1 depending on the degradation function. The coefficient $\beta_0$ will be a constant of $8.889\times 10^{-2}$. The chosen smooth fracture energy will determine which of the $\beta_1, \beta_2,$ or $\beta_3$ coefficients are equal to  $8.889\times 10^{-1}$, the rest being set to zero. Finally, since only one crack driving state function was used, $\beta_4 = 1.778\times 10^1$.

\begin{table}[h!]
    \centering
    \begin{tabular}{c|c}
         Coefficient &  Potential Value \\ \hline
         $\alpha_1, \alpha_2$ & \{0,1\} \\
         $\beta_0$ & $8.889\times 10^{-2}$ \\
         $\beta_1,\beta_2,\beta_3$ & \{0,$8.889\times10^{-1}$\} \\
         $\beta_4$ & $1.778\times10^{1}$
    \end{tabular}
    \caption{Summary of the coefficients used in the ground truth forward problem and targets of the inference.}
    \label{tab:coefficient_summary}
\end{table}
\section{Inference results}
\label{sec:inferenceresults}
\subsection{Sampling from regimes of response}
\label{sec:regimes_of_interest}

The data for inference was obtained from the first three regimes of the load-strain curve: ``elastic'' deformation: A-B, damage initiation: B-C, and damage propagation: C-D. The forward computations were run in serial, with the following being relevant to problem size: The number of rows in the $\boldsymbol{\Xi}_1,\boldsymbol{\Xi}_2$ matrices is the product of the number of degrees of freedom and the number of time steps. For the mesh in \ref{fig:example1_schematic}, with quadratic triangular basis functions for displacements and linear triangular basis functions for damage, there are 217,951 degrees of freedom. The number of time steps in a simulation is  $\sim 1000$ with each region containing $\sim 100$. This scales to a large number of operations with the pseudoinverse:

\begin{equation}
    \boldsymbol{\Theta} = (\boldsymbol{\Xi}^T\boldsymbol{\Xi})^{-1}\boldsymbol{\Xi}^T b
    \label{eq:pseudoinverse}
\end{equation}

Therefore, the largest number of time steps included for inference was limited to 200. In regimes with more than 200 time steps, the last 200 steps were used for inference. We note that 200 time steps is typically larger than the temporal resolution of an experiment.

\subsubsection{Inference of the degradation function}

The results of inferring the degradation function subject to the constraint  $\sum_i \alpha_i = 1$, \ref{eq:linear_vsi_first_min},  appear in \ref{tab:first_inference-full_time}, with values below $1\times 10^{-6}$ set to zero. For each dataset, the coefficients were correctly inferred in all regions of the load-strain curve, accounting for values $\sim 10^{-6}$ being vanishingly small.

\begin{table}[ht!]
\centering
\begin{tabular}{ll|cc|cc}
 Region & $\hat{f}(d)$ & \multicolumn{2}{c|}{$g_1(d)$, Quadratic} & \multicolumn{2}{c}{$g_2(d)$, Quasi-Quadratic} \\
 & & $\alpha_1$ & $\alpha_2$ & $\alpha_1$ & $\alpha_2$ \\ \hline
 & Ground truth & 1 & 0 & 0 & 1 \\ \hline
\multirow{3}{*}{`El.' Def.} & Linear, $f_1(d)$ & 1.0 & 0.0 & $3.028\times 10^{-6}$ & 1.0 \\
 & Single Well, $f_2(d)$ & 1.0 & 0.0 & 0.0 & 1.0 \\
 & Double Well, $f_3(d)$ & 1.0 & 0.0 & 0.0 & 1.0 \\ \hline
\multirow{3}{*}{Dam. Init.} & Linear, $f_1(d)$ & 1.0 & 0.0 & $1.466\times 10^{-6}$ & 1.0 \\
 & Single Well, $f_2(d)$ & 1.0 & 0.0 & 0.0 & 1.0 \\
 & Double Well, $f_3(d)$ & 1.0 & 0.0 & 0.0 & 1.0 \\ \hline
\multirow{3}{*}{Dam. Prop.} & Linear, $f_1(d)$ & 1.0 & 0.0 & $1.465\times 10^{-6}$ & 1.0 \\
 & Single Well, $f_2(d)$ & 1.0 & 0.0 & 0.0 & 1.0 \\
 & Double Well, $f_3(d)$ & 1.0 & 0.0 & 0.0 & 1.0
\end{tabular}
\caption{Inference of the degradation function for different smooth fracture energy functions, $f_j(d)$, using the last 200 time steps of each region of the load-strain curve.}
\label{tab:first_inference-full_time}
\end{table}

For the quadratic degradation function, the correct coefficients were inferred for all three regions of the load-strain curve and for all smooth fracture energy functions. The quasi-quadratic degradation function results were similar, except for the incorrect, but small coefficient for the linear smooth fracture energy in all three regions of the load-strain curve. This is a consequence of the constraint on the $\alpha$ parameters not being exactly satisfied. As mentioned in \cref{sec:regions_of_stressStrain_curve}, this combination of degradation and smooth fracture energy functions is not ideal.

\subsubsection{Inference of smooth fracture energy}

The inferred $\alpha$ coefficients, are used when identifying the smooth fracture energy and crack driving state function. Any $\beta_j < 10^{-6}$ have been set to zero in \ref{tab:second_inference-full_time-linear}, \ref{tab:second_inference-full_time-single_well}, \ref{tab:second_inference-full_time-double_well}. 

Two of the inferred coefficients, $\beta_0$ and $\beta_4$, are not used to identify the smooth fracture energy or the crack driving functions. As shown in \ref{eq:damage_residual_constants}, $\beta_0$ is the re-scaled viscosity and $\beta_4$ represents the crack driving state function; however, since only one functional form was considered for the latter, there are no other coefficients (i.e. $\beta_5,\beta_6,\dots$) to be eliminated. For the smooth fracture energy, we infer the coefficients $\beta_1,\beta_2,$ and $\beta_3$ and expect only one non-zero coefficient in each simulation result (i.e. in each row). The ground truth values for each coefficient are shown in the bottom row of each table and are discussed in \cref{sec:inference_coefficient_summary}.

\begin{table}[ht!]
\centering
\begin{tabular}{ C{5.38em} C{5.18em} | ccccc }
Degradation Function & Region & \multicolumn{1}{c}{$\beta_0$} & \multicolumn{1}{c}{$\beta_1$} & \multicolumn{1}{c}{$\beta_2$} & \multicolumn{1}{c}{$\beta_3$} & $\beta_4$ \\ \hline
\multirow{3}{5.38em}{Quadratic, $g_1(d)$} & `El.' Def. & $8.889\times 10^{-2}$ & $8.889\times 10^{-1}$ & 0.0 & 0.0 & $1.778\times 10^{1}$ \\
 & Dam. Init. & $8.889\times 10^{-2}$ & $8.889\times 10^{-1}$ & 0.0 & 0.0 & $1.778\times 10^{1}$ \\
 & Dam. Prop. & $8.889\times 10^{-2}$ & $8.889\times 10^{-1}$ & 0.0 & 0.0 & $1.778\times 10^{1}$ \\ \hline
\multirow{3}{5.38em}{Quasi-Quadratic, $g_2(d)$} & `El.' Def. & $8.889\times 10^{-2}$ & $8.889\times 10^{-1}$ & $2.945\times 10^{-5}$ & $-2.953\times 10^{-5}$ & $1.778\times 10^{1}$ \\
 & Dam. Init. & $8.889\times 10^{-2}$ & $8.889\times 10^{-1}$ & 0.0 & 0.0 & $1.778\times 10^{1}$ \\
 & Dam. Prop. & $8.889\times 10^{-2}$ & $8.889\times 10^{-1}$ & 0.0 & 0.0 & $1.778\times 10^{1}$ \\ \hline
 & Ground truth & \multicolumn{1}{c}{$8.889\times 10^{-2}$} & \multicolumn{1}{c}{$8.889\times 10^{-1}$} & \multicolumn{1}{c}{0.0} & \multicolumn{1}{c}{0.0} & $1.778\times 10^{1}$ 
\end{tabular}
\caption{Inference of smooth fracture energy and crack driving state function using last 200 time steps of each region if more than 200 time steps exist in the region. Results are shown for each region for both degradation functions with the linear  fracture energy function.}
\label{tab:second_inference-full_time-linear}
\end{table}

\begin{table}[ht!]
\centering
\begin{tabular}{ C{5.4em} C{5.3em} | ccccc }
Degradation Function & Region & \multicolumn{1}{c}{$\beta_0$} & \multicolumn{1}{c}{$\beta_1$} & \multicolumn{1}{c}{$\beta_2$} & \multicolumn{1}{c}{$\beta_3$} & $\beta_4$ \\ \hline 
\multirow{3}{5.4em}{Quadratic, $g_1(d)$} & `El.' Def. & $8.889\times 10^{-2}$ & 0.0 & $8.889\times 10^{-1}$ & 0.0 & $1.778\times 10^{1}$ \\
& Dam. Init. & $8.889\times 10^{-2}$ & 0.0 & $8.889\times 10^{-1}$ & 0.0 & $1.778\times 10^{1}$ \\
& Dam. Prop. & $8.889\times 10^{-2}$ & 0.0 & $8.889\times 10^{-1}$ & 0.0 & $1.778\times 10^{1}$ \\ \hline
\multirow{3}{5.4em}{Quasi- Quadratic, $g_2(d)$} & `El.' Def. & $8.889\times 10^{-2}$ & 0.0 & $8.889\times 10^{-1}$ & $4.554\times 10^{-6}$ & $1.778\times 10^{1}$\\
& Dam. Init. & $8.889\times 10^{-2}$ & 0.0 & $8.889\times 10^{-1}$ & 0.0 & $1.778\times 10^{1}$  \\
& Dam. Prop. & $8.889\times 10^{-2}$ & 0.0 & $8.889\times 10^{-1}$ & 0.0 & $1.778\times 10^{1}$ \\ \hline
& Ground truth & \multicolumn{1}{c}{$8.889\times 10^{-2}$} & \multicolumn{1}{c}{0.0} & \multicolumn{1}{c}{$8.889\times 10^{-1}$} & \multicolumn{1}{c}{0.0} & $1.778\times 10^{1}$
\end{tabular}
\caption{Inference of smooth fracture energy and crack driving state function using last 200 time steps of each region if more than 200 time steps exist in the region. Results are shown for each region for both degradation functions all the single well fracture energy function.}
\label{tab:second_inference-full_time-single_well}
\end{table}

\begin{table}[ht!]
\centering
\begin{tabular}{ C{5.4em} C{5.3em} | ccccc }
Degradation Function & Region & \multicolumn{1}{c}{$\beta_0$} & \multicolumn{1}{c}{$\beta_1$} & \multicolumn{1}{c}{$\beta_2$} & \multicolumn{1}{c}{$\beta_3$} & $\beta_4$ \\ \hline
\multirow{3}{5.4em}{Quadratic, $g_1(d)$} & `El.' Def. & $8.889\times 10^{-2}$ & 0.0 & 0.0 & $8.889\times 10^{-1}$ & $1.778\times 10^{1}$ \\
& Dam. Init. & $8.889\times 10^{-2}$ & 0.0 & 0.0 & $8.889\times 10^{-1}$ & $1.778\times 10^{1}$ \\
& Dam. Prop. & $8.889\times 10^{-2}$ & 0.0 & 0.0 & $8.889\times 10^{-1}$ & $1.778\times 10^{1}$ \\ \hline
\multirow{3}{5.4em}{Quasi- Quadratic, $g_2(d)$} & `El.' Def. & $8.889\times 10^{-2}$ & 0.0 & $-4.574\times 10^{-6}$ & $8.889\times 10^{-1}$ & $1.778\times 10^{1}$ \\
& Dam. Init. & $8.889\times 10^{-2}$ & 0.0 & 0.0 & $8.889\times 10^{-1}$ & $1.778\times 10^{1}$ \\
& Dam. Prop. & $8.889\times 10^{-2}$ & 0.0 & 0.0 & $8.889\times 10^{-1}$ & $1.778\times 10^{1}$ \\ \hline
& Ground truth & \multicolumn{1}{c}{$8.889\times 10^{-2}$} & \multicolumn{1}{c}{0.0} & \multicolumn{1}{c}{0.0} & \multicolumn{1}{c}{$8.889\times 10^{-1}$} & $1.778\times 10^{1}$
\end{tabular}
\caption{Inference of smooth fracture energy and crack driving state function using last 200 time steps of each region if more than 200 time steps exist in the region. Results are shown for each region for both degradation functions with the double well fracture energy function.}
\label{tab:second_inference-full_time-double_well}
\end{table}

The correct parameters/coefficients were determined for all smooth energy functions, in all three regions of the strain-strain curve, with the quadratic degradation function. This is shown in the top three rows of \ref{tab:second_inference-full_time-linear},\ref{tab:second_inference-full_time-single_well}, and \ref{tab:second_inference-full_time-double_well}. In the bottom three rows of each table, with the quasi-quadratic degradation function, the sparsity of the inferred coefficients was reduced in the ``Elastic'' Deformation region of the stress-strain curve. The single and double well smooth fracture energy functions were not properly distinguished by the inference framework. However, this failure of inference was only observed in the ``elastic'' regime, as the smooth fracture energy was successfully identified in the damage initiation and damage propagation regimes for the quasi-quadratic degradation function in all cases.


\subsection{Subsampling time steps}
\label{sec:subsampling_time}

\begin{figure}[ht!]
    \centering
    \includegraphics[width=0.9\linewidth]{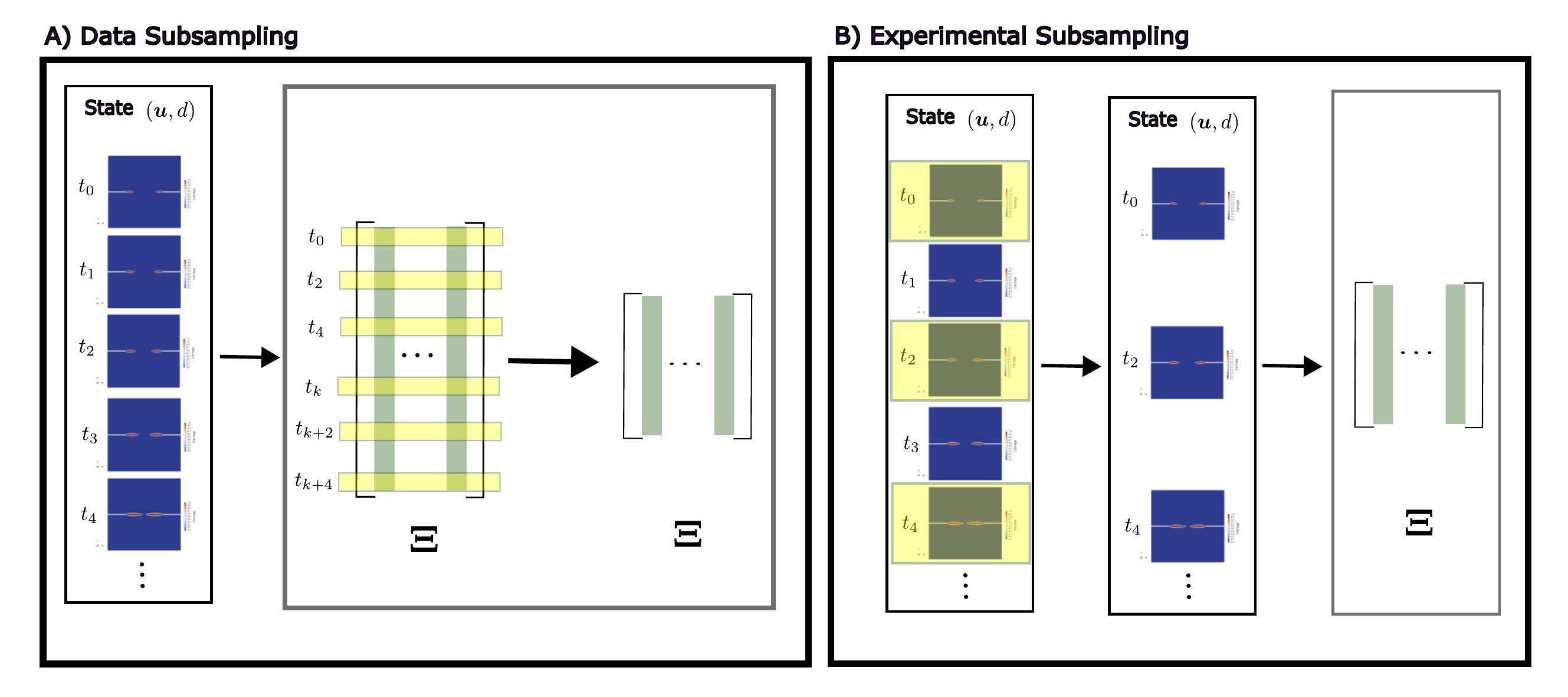}
    \caption{Representation of the subsampling schemes explored. A) Data subsampling where the full-field high-fidelity temporal input data is used to determine the input matrices for inference and the rows of that matrix are then subsampled. B) Experimental subsampling where the temporal input data is directly subsampled and used to evaluate the input matrices for the inference problem.}
    \label{fig:subsampling_schematic}
\end{figure}

While the  results in \cref{sec:regimes_of_interest} used a reasonably dense sampling of data in time, a typical experiment may  provide sparser sampling. We therefore consider the effect of uniform subsampling over time. One approach amounts to directly subsampling from the $\boldsymbol{\Xi}_1$, $\boldsymbol{\Xi}_2$, and $\boldsymbol{b}_2$ matrices. This ``data'' subsampling is shown in panel (A) of \ref{fig:subsampling_schematic}, where the rows of the assembled residual vectors corresponding to the desired time steps (i.e. $t_0, t_2, t_4, \dots$ for subsampling every 2 steps) are extracted to create the new $\boldsymbol{\Xi}$ matrix for the inference problem. \ref{eq:linear_vsi_first_matrices} and \ref{eq:linear_vsi_second_matrices} show that the temporal resolution of the original forward simulation data is preserved with data subsampling. Time derivatives are  determined using the original data with time step $t_{k+1}-t_k = \Delta t$. In the second approach, ``experimental'' subsampling, the input displacement and damage fields are subsampled over time before forming the matrices as seen in panel (B) of \ref{fig:subsampling_schematic}. With experimental subsampling, time derivatives are computed with a larger time step i.e. and $t_{k+2} - t_k = 2\Delta t$.

\subsubsection{Data subsampling results}

\begin{figure}[ht!]
    \centering
    \includegraphics[width=0.85\linewidth]{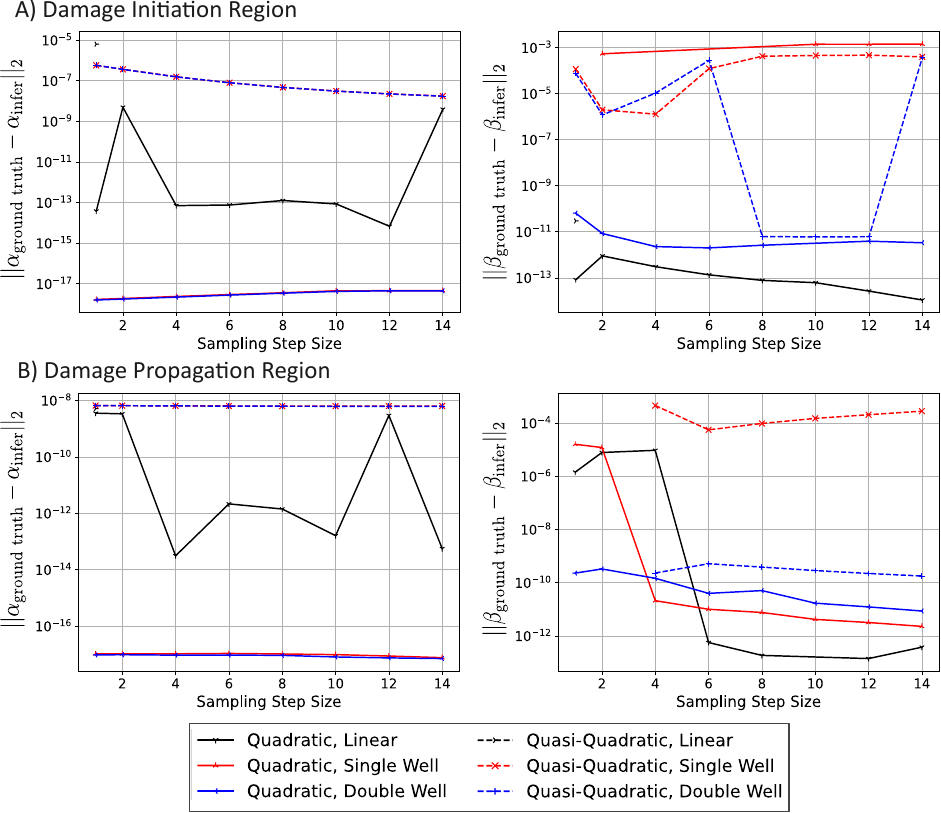}
    \caption{Data subsampling error in the inferred coefficients with increasing sampling interval. Results for  inference of the degradation function (left) and the smooth fracture energy function (right) in the damage initiation (top) and damage propagation (bottom) regions.}
    \label{fig:subsampling_results}
\end{figure}

All forward simulations used a time step of $1\time 10^{-4}$ except for the quasi-quadratic, linear case which used a time step of $5\time 10^{-6}$. \ref{fig:subsampling_results} shows the results of data subsampling as the error of the inferred coefficients against temporal subsampling interval. The error is calculated as the Euclidean norm between the ground truth coefficients from the forward simulation and the inferred coefficients, shown in \ref{eq:inference_error}. These results are shown for the damage initiation and propagation regimes, over which the damage might be visible in an experimental setting.

\begin{equation}
    ||\Theta_{\textrm{ground\ truth}} - \Theta_{\textrm{infer}}||_2
    \label{eq:inference_error}
\end{equation}

For inferring the degradation function, the left side of \ref{fig:subsampling_results}, the error between the exact and inferred coefficients is small (below $1\times 10^{-6}$) and did not significantly change with increased subsampling intervals in both regimes. Similarly on the right side of the Figure, the non-zero coefficients, the phase field functions used in the forward simulation as well as $\beta_0$ and $\beta_4$, were identified in all cases with all subsampling step sizes. However, for this second inference problem, the error decreased for a few cases with increased subsampling intervals, the quadratic linear and single well cases in the damage propagation region. Overall, both the inference of degradation and smooth fracture energy functions were not significantly affected when subjected to ``data'' subsampling.

\subsubsection{Experimental subsampling results}

\begin{figure}[ht!]
    \centering
    \includegraphics[width=0.85\linewidth]{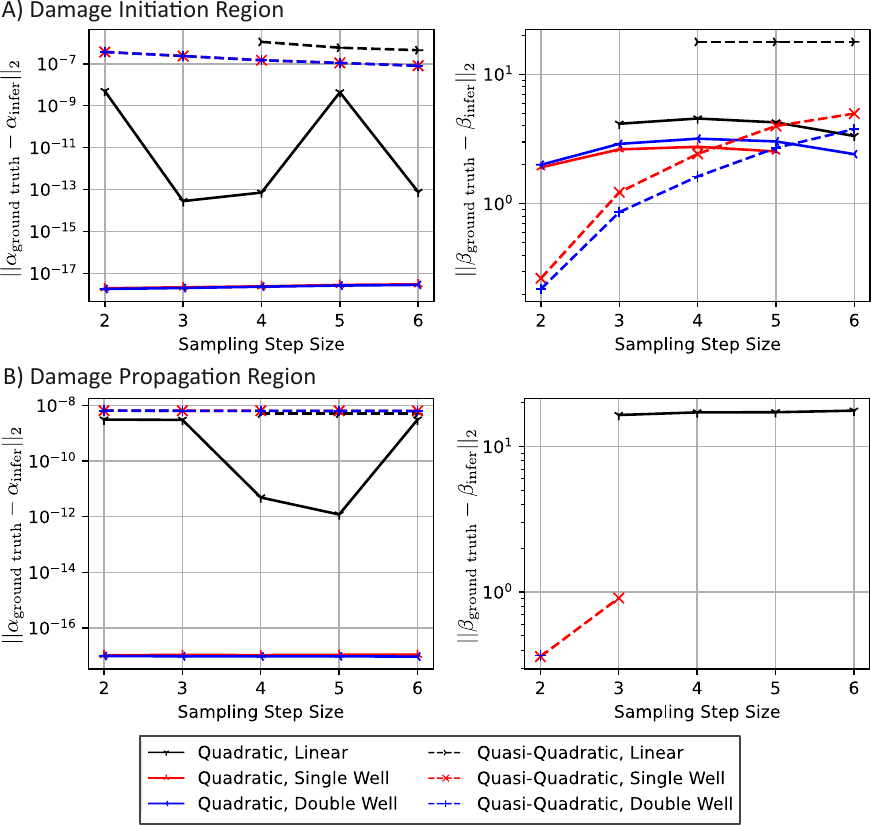}
    \caption{Experimental subsampling over time error in the inferred coefficients with increasing sampling interval. Results for the inference of the degradation function (left) and the smooth fracture energy function (right) in the damage initiation (top) and damage propagation (bottom) regions.}
    \label{fig:experimental_subsampling_results}
\end{figure}

This type of subsampling, which more relevant to experimentally gathered data, did not significantly affect the inference of the degradation function. The error between ground truth and inferred coefficients, \ref{eq:inference_error}, across all cases, was close to machine precision and did not significantly change with increased subsampling interval for both the damage initiation and propagation regions. This is shown on the left side plots in \ref{fig:experimental_subsampling_results}. 

The right side of the figure shows the error in inference of the coefficients of the damage residual, \ref{eq:linear_vsi_second_min}. The missing data points on the bottom right panel in \ref{fig:experimental_subsampling_results} are for subsampling intervals sizes where the solver failed to converge to a solution. This occurred frequently for the inference problem of the damage residual \ref{eq:linear_vsi_second_min} in the damage propagation region. For this problem, the derivative of the damage field ($(d_n - d_{n-1})/\Delta t$) will be affected by the subsampling approach, and the convergence failure is likely due to the higher values for damage in the region, resulting in a larger error in the time derivative with increasing subsampling intervals.  The error in this derivative term is significant, as shown by the inference error on the order of 10 to 100 on the right side of \ref{fig:experimental_subsampling_results}. In the damage initiation region, however, the inference of the smooth fracture energy with a quadratic degradation function was not significantly affected, while the error with a quasi-quadratic degradation function did increase with the  subsampling interval. 

\subsection{Inference from noisy data}

The previous analysis for experimental subsampling was performed on ideal data from full-field forward simulations. Data collected from an experiment will be susceptible to measurement errors which can be modeled as noise. In this analysis, we explore the effect of noise on the inference of coefficients. The applied noise is drawn from a Gaussian distribution with zero mean and is added to each degree of freedom independently and identically distributed within each time step. Guided by the better performance of inference for the quadratic, single-well simulation data with experimental subsampling, we consider this case in the damage initiation and propagation regions, where damage would be visible experimentally, in the neighborhood of a propagating crack. 

\begin{figure}[ht!]
    \centering
    \includegraphics[width=\linewidth]{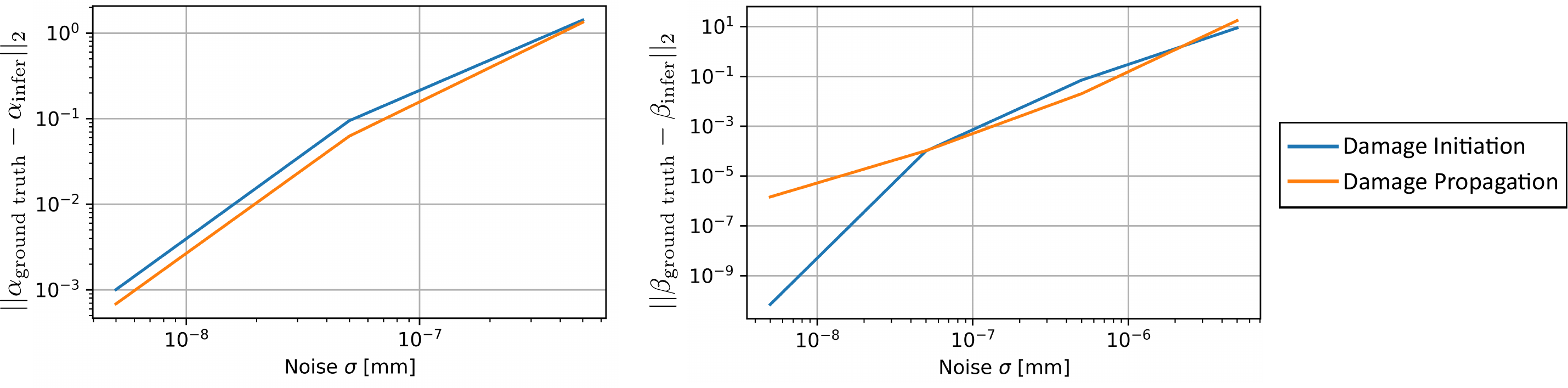}
    \caption{Error between the exact and inferred coefficients with noise added to the vertical displacement, $u_2$. Drawn from a Gaussian distribution with mean of 0 and increasing standard deviation $\sigma$. Results are shown for the quadratic single well problem in the damage initiation and propagation regions. }
    \label{fig:inference_results_noise-quad_single_well}
\end{figure}
 
 \ref{fig:inference_results_noise-quad_single_well} shows the error in the inferred coefficients when noise is applied to the vertical displacement, $u_2$, with increasing standard deviation $\sigma$. At a standard deviation of $5\times 10^{-7}$ mm, the inference fails to identify the correct degradation function associated with the elasticity problem. This is shown on the left side of \ref{fig:inference_results_noise-quad_single_well}, where the error in the inferred coefficients reaches 1.41. The  inference of the smooth fracture energy and coefficients in the damage evolution, is more robust. It fails to identify the correct coefficients at a standard deviation of $5\times 10^{-6}$ mm where the error in the coefficients reaches 8.81 in the damage initiation region and 17.32 in the damage propagation region, as shown on the right side of \ref{fig:inference_results_noise-quad_single_well}. The results for noise applied to the horizontal displacement, $u_1$ and damage phase field $d$ are the same: the correct model can be identified with a standard deviation of $\sigma<5\times 10^{-7}$ mm for the elasticity and $\sigma<5\times 10^{-6}$ mm for the damage problems.

It is important to note that these results correspond to relatively small measurement error. With a Gaussian distribution, a standard deviation of $\sigma=5\times 10^{-7}$ mm leads to a maximum noise value of about $2\times 10^{-6}$ mm. A threshold can be applied to the noise drawn from the distribution so that only those  degrees of freedom with noise above the threshold would be altered. These degrees of freedom would have the largest noise values from the distribution. For example, if the threshold is set to $2\sigma$, only the highest 5\% of noise values would be applied to the data (since we are drawing from a Gaussian distribution). For the mesh used in this study with quadratic triangular elements, this cutoff leaves roughly 10,000 degrees of freedom out of the original 217,951. Implementing this threshold does not change the results shown in \ref{fig:inference_results_noise-quad_single_well}, indicating that a measurement error of $1\times 10^{-6}$ mm, on just a few degrees of freedom, is sufficient to prevent the inference from identifying the correct coefficients in the elasticity problem and an error of $1\times 10^{-5}$ mm is enough to effect the inference of the damage evolution.

The presented results were obtained with the default solver for constraint based optimization from the \texttt{CVXPY} package. The \texttt{SLSQP} method from the \texttt{SciPy} optimization package may also be used. However, this method resulted in slightly higher errors between the inferred and ground truth coefficients. The \texttt{SLSQP} solver was able to identify coefficients with larger values of noise, for example $\sigma=5\times 10^{-5}$ mm, where the \texttt{CVXPY} solver failed to converge. However, the error in inferred coefficients remained large.


\section{Discussion}
\label{sec:discussion}
We have presented variational system identification  for  the degradation function, smooth crack energy function, and history update approach used in phase field fracture models. Inferring the model in the weak form is particularly useful for phase field fracture models since they are developed by minimizing the total energy functional, also leading to the weak form. While identifying the terms within the best fit model, this approach also identifies the parameters used to model damage, the crack length $\ell$ and the critical energy release rate $g_c$. The method presented here enables the selection of the ideal phase field fracture model from data, without prior assumptions about the type of fracture one might expect from a material.


The array of forward simulations generated to verify this method was limited based on restricting the computational expense of generating high-fidelity spatio-temporal data while using functions that are common in the literature. Only two degradation functions that were common in the literature were explored. All three common variants of the smooth fracture energy function from the literature were considered in the forward and inference problems. Finally, only the most common crack driving state function was considered: the critical energy form, which is a subset of the full elastic energy. The stress based criterion, which has also been used in the literature, was not considered because it is not obtained from variational arguments. (It also required significant tuning to get forward simulations to converge.) While the scope of functions used in the forward simulation may appear limited, we note that in this study, the forward data is only required in order to verify the inference method and the results did not vary significantly depending on the choice of degradation or smooth fracture energy functions. However, the combination of  these functions is significant, as shown with the stress-strain curve of the linear, quasi-quadratic simulation. The inference framework presented allows the exploration of a variety of functions one might not typically consider combining. Differences between the selected functions are present in both the full-field solutions and integrated form in the stress-strain curve. The inference approach is able to identify   differences between  functions, such as the single and double well smooth fracture energy functions, with limited data. 

The implementation in this study was performed on plane strain data. However, it should easily extend to three dimensions. The main limitation when extending to high dimensional data is the computational expense. On average, the forward simulations took about five and a half days to run on a 32 cores of a single compute node (AMD EPYC 7742). The inverse problem took approximately five hours to run depending on the number of time steps included in the analysis. The code in this study is serial and used 64 GB of memory on the node.

The benchmark data was generated for a single case of uniaxial tension. More complex loading conditions would significantly affect the compressive/tensile energy split while also introducing shear stresses and strains, impacting the effect of the selected functions on the phase field model. In future work, different geometries and loading scenarios resulting in different crack propagation patterns should be explored. This work also considered a single  material model. The nonlinear elasticity model introduced by the Saint-Venant Kirchoff model was limited and the inference approach should be verified with more advanced models for nonlinear elasticity such as quadratic-logarithmic, or neo-Hookean and HGO models for soft materials.


The St. Venant-Kirchhoff and phase field fracture models in this first communication have parameters that appear as linear coefficients of different functions or operators. There is a possibility to identify more complex degradation and geometric functions than exist in the literature by considering combinations of functions that would introduce nonlinear dependencies on the parameters, requiring nonlinear regression. With such an extension in mind, the code used here does indeed employ nonlinear optimization, notwithstanding its greater computational expense due to the repeated calculation of the objective and gradient functions over all time steps. 

Within the inference framework, standard regression techniques were used without considering operator elimination. In the early work on SINDy, sparse regression was used to identify the dominant operators in the system.\cite{messenger_weak_2021} In Wang et al., stepwise regression was used to eliminate operators in an iterative scheme in order to identify the governing equation.\cite{wang_perspective_2020} In this work, a relatively small library of terms was compared. This rarely led to difficulties when identifying the governing model because the full-field synthetic data proved to be sufficiently informative  to distinguish between and infer from the smaller collection of terms.


There are a few considerations when planning to apply this method with experimental data. First, methods like Digital Image Correlation (DIC) can be used to estimate the displacement field. However, the damage field needs to be extracted from the neighborhood of cracks in experimental samples--perhaps using optical or other visualization techniques. Its normalization to a phase field would then be a simple step. Depending on the quality of the experimental setup, this data can provide measurements with high accuracy and low noise. However, the temporal resolution may be limited, so it is important to ensure that the inference approach works with sparse temporal data. Utilizing the data subsampling method, the inference was still able to identify the governing phase-field fracture model. However, experimental subsampling, as may be expected, led to a deterioration of the time derivative,  affecting the success of the inference. If using coarse time data, one should expect this type of error and adjust the evaluation of the derivative accordingly, using more robust methods to estimate any rate term.

The effect of noise on system identification is significant. This is confirmed in our study, with relatively small amounts of noise impacting the ability to infer the governing model. The value of noise required to stop correctly identifying the governing model was inspected by applying a threshold to the noise distribution, showing that a few degrees of freedom with larger values of noise are sufficient to effect the ability to infer the model. Further study of the effect of noise on a coarser mesh would be useful, since the effects of noise on larger elements should be smaller. In this regard, we note that noise levels of $\sim 10^{-6}$ mm, at which inference was successful for the phase field fracture model, are  of the order of accuracy of the DIC techniques for full-field displacemnt measurements. Therefore, the inference approach presented here is indeed viable. Inference of elasticity parameters failed at noise $\sim 10^{-7}$ mm, but other approaches exist for high-precision inference of elasticity parameters. The variational system identification approach should be reserved for the phase field fracture problem, instead.

Throughout this analysis, the first inference problem which identifies the degradation function affecting elasticity, was susceptible to smaller degrees of noise than the second inference problem, identifying the smooth fracture energy in the damage evolution problem. This is due to the higher degree of nonlinearity in the elasticity model, where the effects of noise are exacerbated. Bayesian inference can be combined with variational system identification to provide uncertainty quantification via repeated inference of the coefficients.\cite{wang_perspective_2020}  PDE constrained optimization, evaluating the gradient with the adjoint method, has been found to improve parameter estimation following model identification.\cite{wang_inference_2021-1}

As a closing remark we note that  varying degrees of success were obtained for inference with different phase field fracture models. This does not imply a recommendation for one form over the other. Instead, the computational modeler or experimentalist should be aware of the aspects of identifiability of these models while inferring them from data--as well as while using them.


\appendix

\section{Derivation of the Euler-Lagrange equations}
\label{appendix:euler_lagrange}

\subsection{Variation in displacement}

The variational derivative of $\Psi$ with respect to $u$, using the variation $\varepsilon$ and weighting functions $w_u$:
\begin{equation}
\begin{split}
    \delta\Psi(u,w_u) &= \dv{\varepsilon}\left.\Psi(u+\varepsilon w_u)\right\vert_{\varepsilon=0}\\
    &= \dv{\varepsilon} \left.\left\{\int_\Omega g(d)\psi_e(\boldsymbol{F}(u+\varepsilon w_u)) d\Omega + \int_\Omega g_c \gamma_\ell(d,\nabla d)d\Omega \right\}\right\vert_{\varepsilon=0} \\
    &= \left.\int_\Omega g(d) \dv{\varepsilon} \psi_e(\boldsymbol{F}(u+\varepsilon w_u)) d\Omega + \cancelto{0}{\dv{\varepsilon}\int_\Omega g_c \gamma_\ell(d,\nabla d)d\Omega} \hspace{1.5em} \right\vert_{\varepsilon=0} \\
    &= \left.\int_\Omega g(d) \fdv{\varepsilon} \psi_e(\boldsymbol{F}(u+\varepsilon w_u)) d\Omega \right\vert_{\varepsilon=0}
\end{split}
\end{equation}

Note that we need to apply the chain rule and evaluate the derivatives of the tensor $\boldsymbol{F}$
\begin{equation}
    \begin{split}
        \dv{\varepsilon}\psi_e(\boldsymbol{F}(u + \varepsilon w_u)) &= \dv{\psi_e(\boldsymbol{F}(u+\varepsilon w_u))}{\boldsymbol{F}(u)}:\dv{\boldsymbol{F}(u+\varepsilon w_u)}{\varepsilon} \\
        &= \dv{\psi_e(\boldsymbol{F}(u+\varepsilon w_u))}{\boldsymbol{F}(u)}:\dv{\varepsilon}\left(\boldsymbol{I} + \dv{x}(u+\varepsilon w_u)\right) \\
        &= \dv{\psi_e(\boldsymbol{F}(u+\varepsilon w_u))}{\boldsymbol{F}(u)}:\dv{\varepsilon}\boldsymbol{I} + \dv{x}\dv{\varepsilon}(u+\varepsilon w_u) \\
        &= \dv{\psi_e(\boldsymbol{F}(u+\varepsilon w_u))}{\boldsymbol{F}(u)}:\dv{w_u}{x}
    \end{split}
\end{equation}

Plugging this back into our variational derivative and considering the variation equal to zero, $\varepsilon=0$, we have:
\begin{equation}
    \begin{split}
        \delta\Psi(u,w_u) &= \left.\int_\Omega g(d) \dv{\psi_e(\boldsymbol{F}(u+\varepsilon w_u))}{\boldsymbol{F}(u)}:\dv{w_u}{x} d\Omega \right\vert_{\varepsilon=0} \\
        &= \int_\Omega g(d) \dv{\psi_e(\boldsymbol{F}(u)}{\boldsymbol{F}(u)}:\nabla w_u d\Omega
    \end{split}
\end{equation}

\subsection{Variation in damage}

Now, the variational derivative of $\Psi$ with respect to $d$, using the variation $\varepsilon$ and weighting functions $w_d$ and plugging in the form of regularized crack energy $\gamma_\ell$ from \ref{eq:energy_density}:

\begin{equation}
\begin{split}
    \delta\Psi(d,w_d) &= \dv{\varepsilon}\left.\Psi(d_0+\varepsilon w_d)\right\vert_{\varepsilon=0} \\
    &= \dv{\varepsilon}\left. \int_\Omega g(d+\varepsilon w_d)\psi_e(\boldsymbol{F}(u)) + g_c\gamma_\ell(d+\varepsilon w_d, \nabla(d+\varepsilon w_d)) d\Omega \right\vert_{\varepsilon=0} \\
    &= \dv{\varepsilon} \left.\left\{ \int_\Omega g(d+\varepsilon w_d) \psi_e(\boldsymbol{F}(u)) d\Omega + \int_\Omega g_c c_s (\beta f(d+\varepsilon w_d) + \frac{1}{2}\abs{\nabla (d+\varepsilon w_d)}^2 ) d\Omega \right\}\right\vert_{\varepsilon=0} \\
    &= \left. \int_\Omega \dv{\varepsilon} g(d+\varepsilon w_d) \psi_e(\boldsymbol{F}(u)) d\Omega + \int_\Omega g_c c_s \beta \dv{\varepsilon}f(d+\varepsilon w_d) d\Omega + \int_\Omega g_c\kappa \frac{1}{2} \dv{\varepsilon} \abs{ \nabla (d+\varepsilon w_d)}^2 d\Omega \right\vert_{\varepsilon=0}
\end{split}
\end{equation}

For the two terms, once again, we apply the chain rule. Shown belof for function $g(d)$, but equivalent for the function $f(d)$:
\begin{equation}
    \begin{split}
        \dv{\varepsilon}g(d+\varepsilon w_d) &= \dv{g(d+\varepsilon w_d)}{d}\dv{(d+\varepsilon w_d)}{\varepsilon} \\
        &= \dv{g(d+\varepsilon w_d)}{d}w_d
    \end{split}
\end{equation}

Then, the expansion for the last term:
\begin{equation}
\begin{split}
    \dv{\varepsilon}\frac{1}{2}\abs{\nabla(d+\varepsilon w_d)}^2 &= \dv{\varepsilon}\frac{1}{2} \left(\nabla (d+\varepsilon w_d) \cdot \nabla (d+\varepsilon w_d) \right) \\
    &= \frac{1}{2}\left(\dv{\varepsilon}\nabla d_\varepsilon \cdot \nabla d_\varepsilon + \nabla d_\varepsilon \cdot \dv{\varepsilon}\nabla d_\varepsilon \right)\\
    &= \frac{1}{2}\left(2\nabla w_d \cdot \nabla d_\varepsilon \right) \\ 
    &= \nabla w_d \cdot \nabla (d+\varepsilon w_d)
\end{split}
\end{equation}

Plugging this back in and evaluating at $\varepsilon = 0$:
\begin{equation}
    \begin{split}
        \delta\Psi(d,w_d) &= \left. \int_\Omega \dv{g(d+\varepsilon w_d)}{d}w_d \psi_e(\boldsymbol{F}(u)) d\Omega + \int_\Omega g_c c_s\beta \dv{f(d+\varepsilon w_d)}{d}w_d d\Omega + \int_\Omega g_c c_s \nabla w_d \cdot \nabla (d+\varepsilon w_d) d\Omega \right\vert_{\varepsilon=0} \\
        &= \int_\Omega \dv{g(d)}{d}w_d \psi_e(\boldsymbol{F}(u)) d\Omega + \int_\Omega g_c c_s \beta \dv{f(d)}{d}w_d d\Omega + \int_\Omega g_c c_s \nabla w_d \cdot \nabla d \ d\Omega \\
        &= \int_\Omega \dv{g(d)}{d}w_d \psi_e(\boldsymbol{F}(u)) d\Omega + \int_\Omega g_c c_s \left( \beta \dv{f(d)}{d}w_d + \nabla w_d \cdot \nabla d\right) \ d\Omega 
    \end{split}
    \label{eq:damage_weak_form_steady}
\end{equation}


\newpage
\pagestyle{bibliography}
\bibliography{pff_references,systemID_references}

\begin{thebibliography}{10}
\providecommand{\url}[1]{#1}
\csname url@samestyle\endcsname
\providecommand{\newblock}{\relax}
\providecommand{\bibinfo}[2]{#2}
\providecommand{\BIBentrySTDinterwordspacing}{\spaceskip=0pt\relax}
\providecommand{\BIBentryALTinterwordstretchfactor}{4}
\providecommand{\BIBentryALTinterwordspacing}{\spaceskip=\fontdimen2\font plus
\BIBentryALTinterwordstretchfactor\fontdimen3\font minus
  \fontdimen4\font\relax}
\providecommand{\BIBforeignlanguage}[2]{{%
\expandafter\ifx\csname l@#1\endcsname\relax
\typeout{** WARNING: IEEEtran.bst: No hyphenation pattern has been}%
\typeout{** loaded for the language `#1'. Using the pattern for}%
\typeout{** the default language instead.}%
\else
\language=\csname l@#1\endcsname
\fi
#2}}
\providecommand{\BIBdecl}{\relax}
\BIBdecl

\bibitem{rudraraju2012predictions}
S.~Rudraraju, A.~Salvi, K.~Garikipati, and A.~M. Waas, ``Predictions of crack
  propagation using a variational multiscale approach and its application to
  fracture in laminated fiber reinforced composites,'' \emph{Composite
  structures}, vol.~94, no.~11, pp. 3336--3346, 2012.

\bibitem{francfort_revisiting_1998}
G.~Francfort and J.-J. Marigo, ``Revisiting brittle fracture as an energy
  minimization problem,'' \emph{Journal of the Mechanics and Physics of
  Solids}, vol.~46, no.~8, pp. 1319--1342, Aug. 1998.

\bibitem{griffith_phenomena_1920}
A.~A. Griffith and M.~Eng, ``The {{Phenomena}} of {{Rupture}} and {{Flow}} in
  {{Solids}},'' \emph{Philosophical Transactions of the Royal Society of
  London}, vol. 221, pp. 163--198, 1920.

\bibitem{bourdin_numerical_2000}
B.~Bourdin, G.~Francfort, and J.-J. Marigo, ``Numerical experiments in
  revisited brittle fracture,'' \emph{Journal of the Mechanics and Physics of
  Solids}, vol.~48, no.~4, pp. 797--826, Apr. 2000.

\bibitem{ambrosio_approximation_1990}
L.~Ambrosio and V.~M. Tortorelli, ``Approximation of functional depending on
  jumps by elliptic functional via t-convergence,'' \emph{Communications on
  Pure and Applied Mathematics}, vol.~43, no.~8, pp. 999--1036, Dec. 1990.

\bibitem{amor_regularized_2009}
H.~Amor, J.-J. Marigo, and C.~Maurini, ``Regularized formulation of the
  variational brittle fracture with unilateral contact: {{Numerical}}
  experiments,'' \emph{Journal of the Mechanics and Physics of Solids},
  vol.~57, no.~8, pp. 1209--1229, Aug. 2009.

\bibitem{miehe_phase_2010}
C.~Miehe, M.~Hofacker, and F.~Welschinger, ``A phase field model for
  rate-independent crack propagation: {{Robust}} algorithmic implementation
  based on operator splits,'' \emph{Computer Methods in Applied Mechanics and
  Engineering}, vol. 199, no. 45-48, pp. 2765--2778, Nov. 2010.

\bibitem{kuhn_phase_2008}
C.~Kuhn and R.~M{\"u}ller, ``A phase field model for fracture,'' \emph{PAMM},
  vol.~8, no.~1, pp. 10\,223--10\,224, Dec. 2008.

\bibitem{miehe_phase_2015}
C.~Miehe, L.-M. Sch{\"a}nzel, and H.~Ulmer, ``Phase field modeling of fracture
  in multi-physics problems. {{Part I}}. {{Balance}} of crack surface and
  failure criteria for brittle crack propagation in thermo-elastic solids,''
  \emph{Computer Methods in Applied Mechanics and Engineering}, vol. 294, pp.
  449--485, Sep. 2015.

\bibitem{ambati_review_2015}
M.~Ambati, T.~Gerasimov, and L.~De~Lorenzis, ``A review on phase-field models
  of brittle fracture and a new fast hybrid formulation,'' \emph{Computational
  Mechanics}, vol.~55, no.~2, pp. 383--405, Feb. 2015.

\bibitem{kuhn_degradation_2015}
C.~Kuhn, A.~Schl{\"u}ter, and R.~M{\"u}ller, ``On degradation functions in
  phase field fracture models,'' \emph{Computational Materials Science}, vol.
  108, pp. 374--384, Oct. 2015.

\bibitem{sargado_highaccuracy_2018}
J.~M. Sargado, E.~Keilegavlen, I.~Berre, and J.~M. Nordbotten, ``High-accuracy
  phase-field models for brittle fracture based on a new family of degradation
  functions,'' \emph{Journal of the Mechanics and Physics of Solids}, vol. 111,
  pp. 458--489, Feb. 2018.

\bibitem{pham_gradient_2011}
K.~Pham, H.~Amor, J.-J. Marigo, and C.~Maurini, ``Gradient {{Damage Models}}
  and {{Their Use}} to {{Approximate Brittle Fracture}},'' \emph{International
  Journal of Damage Mechanics}, vol.~20, no.~4, pp. 618--652, May 2011.

\bibitem{wu_unified_2017}
J.-Y. Wu, ``A unified phase-field theory for the mechanics of damage and
  quasi-brittle failure,'' \emph{Journal of the Mechanics and Physics of
  Solids}, vol. 103, pp. 72--99, Jun. 2017.

\bibitem{bourdin_variational_2008}
B.~Bourdin, G.~A. Francfort, and J.-J. Marigo, ``The {{Variational Approach}}
  to {{Fracture}},'' \emph{Journal of Elasticity}, vol.~91, no. 1-3, pp.
  5--148, Apr. 2008.

\bibitem{wu_phasefield_2020}
J.-Y. Wu, V.~P. Nguyen, C.~T. Nguyen, D.~Sutula, S.~Sinaie, and S.~P. Bordas,
  ``Phase-field modeling of fracture,'' in \emph{Advances in {{Applied
  Mechanics}}}.\hskip 1em plus 0.5em minus 0.4em\relax Elsevier, 2020, vol.~53,
  pp. 1--183.

\bibitem{diehl_comparative_2022}
P.~Diehl, R.~Lipton, T.~Wick, and M.~Tyagi, ``A comparative review of
  peridynamics and phase-field models for engineering fracture mechanics,''
  \emph{Computational Mechanics}, vol.~69, no.~6, pp. 1259--1293, Jun. 2022.

\bibitem{freddi_regularized_2010}
F.~Freddi and G.~{Royer-Carfagni}, ``Regularized variational theories of
  fracture: {{A}} unified approach,'' \emph{Journal of the Mechanics and
  Physics of Solids}, vol.~58, no.~8, pp. 1154--1174, Aug. 2010.

\bibitem{suh_phase_2020}
H.~S. Suh, W.~Sun, and D.~T. O'Connor, ``A phase field model for cohesive
  fracture in micropolar continua,'' \emph{Computer Methods in Applied
  Mechanics and Engineering}, vol. 369, p. 113181, Sep. 2020.

\bibitem{alessi_comparison_2018}
R.~Alessi, M.~Ambati, T.~Gerasimov, S.~Vidoli, and L.~De~Lorenzis, ``Comparison
  of {{Phase-Field Models}} of {{Fracture Coupled}} with {{Plasticity}},'' in
  \emph{Advances in {{Computational Plasticity}}}, E.~O{\~n}ate, D.~Peric,
  E.~De~Souza~Neto, and M.~Chiumenti, Eds.\hskip 1em plus 0.5em minus
  0.4em\relax Cham: Springer International Publishing, 2018, vol.~46, pp.
  1--21.

\bibitem{svolos_convexity_2023}
L.~Svolos, J.~N. Plohr, G.~Manzini, and H.~M. Mourad, ``On the convexity of
  phase-field fracture formulations: {{Analytical}} study and comparison of
  various degradation functions,'' \emph{International Journal of Non-Linear
  Mechanics}, vol. 150, p. 104359, Apr. 2023.

\bibitem{deborst_gradient_2016}
R.~De~Borst and C.~V. Verhoosel, ``Gradient damage vs phase-field approaches
  for fracture: {{Similarities}} and differences,'' \emph{Computer Methods in
  Applied Mechanics and Engineering}, vol. 312, pp. 78--94, Dec. 2016.

\bibitem{wells2004discontinuous}
G.~N. Wells, K.~Garikipati, and L.~Molari, ``A discontinuous galerkin
  formulation for a strain gradient-dependent damage model,'' \emph{Computer
  Methods in Applied Mechanics and Engineering}, vol. 193, no. 33-35, pp.
  3633--3645, 2004.

\bibitem{molari2006discontinuous}
L.~Molari, G.~N. Wells, K.~Garikipati, and F.~Ubertini, ``A discontinuous
  galerkin method for strain gradient-dependent damage: study of interpolations
  and convergence,'' \emph{Computer Methods in Applied Mechanics and
  Engineering}, vol. 195, no. 13-16, pp. 1480--1498, 2006.

\bibitem{wu_parameter_2021}
T.~Wu, B.~Rosi{\'c}, L.~De~Lorenzis, and H.~G. Matthies, ``Parameter
  identification for phase-field modeling of fracture: A {{Bayesian}} approach
  with sampling-free update,'' \emph{Computational Mechanics}, vol.~67, no.~2,
  pp. 435--453, Feb. 2021.

\bibitem{kosin_parameter_2024}
V.~Kosin, A.~Fau, C.~Jailin, F.~Hild, and T.~Wick, ``Parameter identification
  of a phase-field fracture model using integrated digital image correlation,''
  \emph{Computer Methods in Applied Mechanics and Engineering}, vol. 420, p.
  116689, Feb. 2024.

\bibitem{lennart_ljung_system_1999}
\BIBentryALTinterwordspacing
{Lennart Ljung}, ``System {Identification},'' in \emph{Signal {Analysis} and
  {Prediction}}, 1999, pp. 263--282. [Online]. Available:
  \url{https://doi.org/10.1007/978-1-4612-1768-8}
\BIBentrySTDinterwordspacing

\bibitem{schmidt_distilling_2009}
\BIBentryALTinterwordspacing
M.~Schmidt and H.~Lipson, ``\BIBforeignlanguage{en}{Distilling {Free}-{Form}
  {Natural} {Laws} from {Experimental} {Data}},''
  \emph{\BIBforeignlanguage{en}{Science}}, vol. 324, no. 5923, pp. 81--85, Apr.
  2009. [Online]. Available:
  \url{https://www.science.org/doi/10.1126/science.1165893}
\BIBentrySTDinterwordspacing

\bibitem{brunton_discovering_2016}
\BIBentryALTinterwordspacing
S.~L. Brunton, J.~L. Proctor, and J.~N. Kutz,
  ``\BIBforeignlanguage{en}{Discovering governing equations from data by sparse
  identification of nonlinear dynamical systems},''
  \emph{\BIBforeignlanguage{en}{Proceedings of the National Academy of
  Sciences}}, vol. 113, no.~15, pp. 3932--3937, Apr. 2016. [Online]. Available:
  \url{http://www.pnas.org/lookup/doi/10.1073/pnas.1517384113}
\BIBentrySTDinterwordspacing

\bibitem{rowley_spectral_2009}
\BIBentryALTinterwordspacing
C.~W. Rowley, I.~Mezić, S.~Bagheri, P.~Schlatter, and D.~S. Henningson,
  ``\BIBforeignlanguage{en}{Spectral analysis of nonlinear flows},''
  \emph{\BIBforeignlanguage{en}{Journal of Fluid Mechanics}}, vol. 641, pp.
  115--127, Dec. 2009. [Online]. Available:
  \url{https://www.cambridge.org/core/product/identifier/S0022112009992059/type/journal_article}
\BIBentrySTDinterwordspacing

\bibitem{gonzalez-garcia_identification_1998}
\BIBentryALTinterwordspacing
R.~González-García, R.~Rico-Martínez, and I.~Kevrekidis,
  ``\BIBforeignlanguage{en}{Identification of distributed parameter systems:
  {A} neural net based approach},'' \emph{\BIBforeignlanguage{en}{Computers \&
  Chemical Engineering}}, vol.~22, pp. S965--S968, Mar. 1998. [Online].
  Available:
  \url{https://linkinghub.elsevier.com/retrieve/pii/S0098135498001914}
\BIBentrySTDinterwordspacing

\bibitem{raissi_physics-informed_2019}
\BIBentryALTinterwordspacing
M.~Raissi, P.~Perdikaris, and G.~Karniadakis,
  ``\BIBforeignlanguage{en}{Physics-informed neural networks: {A} deep learning
  framework for solving forward and inverse problems involving nonlinear
  partial differential equations},'' \emph{\BIBforeignlanguage{en}{Journal of
  Computational Physics}}, vol. 378, pp. 686--707, Feb. 2019. [Online].
  Available:
  \url{https://linkinghub.elsevier.com/retrieve/pii/S0021999118307125}
\BIBentrySTDinterwordspacing

\bibitem{huang_variational_2022}
\BIBentryALTinterwordspacing
S.~Huang, Z.~He, B.~Chem, and C.~Reina, ``\BIBforeignlanguage{en}{Variational
  {Onsager} {Neural} {Networks} ({VONNs}): {A} thermodynamics-based variational
  learning strategy for non-equilibrium {PDEs}},''
  \emph{\BIBforeignlanguage{en}{Journal of the Mechanics and Physics of
  Solids}}, vol. 163, p. 104856, Jun. 2022. [Online]. Available:
  \url{https://linkinghub.elsevier.com/retrieve/pii/S0022509622000692}
\BIBentrySTDinterwordspacing

\bibitem{linka_new_2023}
\BIBentryALTinterwordspacing
K.~Linka and E.~Kuhl, ``\BIBforeignlanguage{en}{A new family of {Constitutive}
  {Artificial} {Neural} {Networks} towards automated model discovery},''
  \emph{\BIBforeignlanguage{en}{Computer Methods in Applied Mechanics and
  Engineering}}, vol. 403, p. 115731, Jan. 2023. [Online]. Available:
  \url{https://linkinghub.elsevier.com/retrieve/pii/S0045782522006867}
\BIBentrySTDinterwordspacing

\bibitem{wang_variational_2019}
Z.~Wang, X.~Huan, and K.~Garikipati, ``\BIBforeignlanguage{en}{Variational
  system identification of the partial differential equations governing the
  physics of pattern-formation: {Inference} under varying fidelity and
  noise},'' Nov. 2019.

\bibitem{wang_perspective_2020}
\BIBentryALTinterwordspacing
Z.~Wang, B.~Wu, K.~Garikipati, and X.~Huan, ``\BIBforeignlanguage{en}{A
  {Perspective} on {Regression} and {Bayesian} {Approaches} for {System}
  {Identification} of {Pattern} {Formation} {Dynamics}},''
  \emph{\BIBforeignlanguage{en}{arXiv:2001.05646 [physics]}}, Mar. 2020, arXiv:
  2001.05646. [Online]. Available: \url{http://arxiv.org/abs/2001.05646}
\BIBentrySTDinterwordspacing

\bibitem{wang_variational_2021}
\BIBentryALTinterwordspacing
Z.~Wang, X.~Huan, and K.~Garikipati, ``\BIBforeignlanguage{en}{Variational
  system identification of the partial differential equations governing
  microstructure evolution in materials: {Inference} over sparse and spatially
  unrelated data},'' \emph{\BIBforeignlanguage{en}{Computer Methods in Applied
  Mechanics and Engineering}}, vol. 377, p. 113706, Apr. 2021. [Online].
  Available:
  \url{https://linkinghub.elsevier.com/retrieve/pii/S0045782521000426}
\BIBentrySTDinterwordspacing

\bibitem{wang_inference_2021}
Z.~Wang, J.~B. Estrada, E.~M. Arruda, and K.~Garikipati,
  ``\BIBforeignlanguage{en}{Inference of deformation mechanisms and
  constitutive response of soft material surrogates of biological tissue by
  full-field characterization and data-driven variational system
  identification},'' 2021.

\bibitem{nikolov2022ogden}
D.~P. Nikolov, S.~Srivastava, B.~A. Abeid, U.~M. Scheven, E.~M. Arruda,
  K.~Garikipati, and J.~B. Estrada, ``Ogden material calibration via magnetic
  resonance cartography, parameter sensitivity and variational system
  identification,'' \emph{Philosophical Transactions of the Royal Society A},
  vol. 380, no. 2234, p. 20210324, 2022.

\bibitem{ho2023oscillatory}
K.~K. Ho, S.~Srivastava, P.~C. Kinnunen, K.~Garikipati, G.~D. Luker, and K.~E.
  Luker, ``Oscillatory erk signaling and morphology determine heterogeneity of
  breast cancer cell chemotaxis via mek-erk and p38-mapk signaling pathways,''
  \emph{Bioengineering}, vol.~10, no.~2, p. 269, 2023.

\bibitem{srivastava_pattern_2024}
\BIBentryALTinterwordspacing
S.~Srivastava and K.~Garikipati, ``\BIBforeignlanguage{en}{Pattern formation in
  dense populations studied by inference of nonlinear diffusion‐reaction
  mechanisms},'' \emph{\BIBforeignlanguage{en}{International Journal for
  Numerical Methods in Engineering}}, p. e7475, Mar. 2024. [Online]. Available:
  \url{https://onlinelibrary.wiley.com/doi/10.1002/nme.7475}
\BIBentrySTDinterwordspacing

\bibitem{kinnunen2024inference}
P.~C. Kinnunen, S.~Srivastava, Z.~Wang, K.~K. Ho, B.~A. Humphries, S.~Chen,
  J.~J. Linderman, G.~D. Luker, K.~E. Luker, and K.~Garikipati, ``Inference of
  weak-form partial differential equations describing migration and
  proliferation mechanisms in wound healing experiments on cancer cells,''
  \emph{ArXiv}, pp. arXiv--2302, 2024.

\bibitem{wang2020system}
Z.~Wang, X.~Zhang, G.~Teichert, M.~Carrasco-Teja, and K.~Garikipati, ``System
  inference for the spatio-temporal evolution of infectious diseases: Michigan
  in the time of covid-19,'' \emph{Computational Mechanics}, vol.~66, pp.
  1153--1176, 2020.

\bibitem{wang2021system}
Z.~Wang, M.~Carrasco-Teja, X.~Zhang, G.~H. Teichert, and K.~Garikipati,
  ``System inference via field inversion for the spatio-temporal progression of
  infectious diseases: Studies of covid-19 in michigan and mexico,''
  \emph{Archives of Computational Methods in Engineering}, vol.~28, pp.
  4283--4295, 2021.

\bibitem{messenger_weak_2021}
\BIBentryALTinterwordspacing
D.~A. Messenger and D.~M. Bortz, ``\BIBforeignlanguage{en}{Weak {SINDy} for
  partial differential equations},'' \emph{\BIBforeignlanguage{en}{Journal of
  Computational Physics}}, vol. 443, p. 110525, Oct. 2021. [Online]. Available:
  \url{https://linkinghub.elsevier.com/retrieve/pii/S0021999121004204}
\BIBentrySTDinterwordspacing

\bibitem{wentz_derivative-based_2023}
\BIBentryALTinterwordspacing
J.~Wentz and A.~Doostan, ``\BIBforeignlanguage{en}{Derivative-based {SINDy}
  ({DSINDy}): {Addressing} the challenge of discovering governing equations
  from noisy data},'' \emph{\BIBforeignlanguage{en}{Computer Methods in Applied
  Mechanics and Engineering}}, vol. 413, p. 116096, Aug. 2023. [Online].
  Available:
  \url{https://linkinghub.elsevier.com/retrieve/pii/S0045782523002207}
\BIBentrySTDinterwordspacing

\bibitem{li_gradient_2016}
T.~Li, J.-J. Marigo, D.~Guilbaud, and S.~Potapov, ``Gradient damage modeling of
  brittle fracture in an explicit dynamics context,'' \emph{International
  Journal for Numerical Methods in Engineering}, vol. 108, no.~11, pp.
  1381--1405, Dec. 2016.

\bibitem{steinke_phasefield_2019}
C.~Steinke and M.~Kaliske, ``A phase-field crack model based on directional
  stress decomposition,'' \emph{Computational Mechanics}, vol.~63, no.~5, pp.
  1019--1046, May 2019.

\bibitem{borden_phasefield_2016}
M.~J. Borden, T.~J. Hughes, C.~M. Landis, A.~Anvari, and I.~J. Lee, ``A
  phase-field formulation for fracture in ductile materials: {{Finite}}
  deformation balance law derivation, plastic degradation, and stress
  triaxiality effects,'' \emph{Computer Methods in Applied Mechanics and
  Engineering}, vol. 312, pp. 130--166, Dec. 2016.

\bibitem{karma_phasefield_2001}
A.~Karma, D.~A. Kessler, and H.~Levine, ``Phase-{{Field Model}} of {{Mode III
  Dynamic Fracture}},'' \emph{Physical Review Letters}, vol.~87, no.~4, p.
  045501, Jul. 2001.

\bibitem{geelen_phasefield_2019}
R.~J. Geelen, Y.~Liu, T.~Hu, M.~R. Tupek, and J.~E. Dolbow, ``A phase-field
  formulation for dynamic cohesive fracture,'' \emph{Computer Methods in
  Applied Mechanics and Engineering}, vol. 348, pp. 680--711, May 2019.

\bibitem{lorentz_modelling_2012}
E.~Lorentz, S.~Cuvilliez, and K.~Kazymyrenko, ``Modelling large crack
  propagation: From gradient damage to cohesive zone models,''
  \emph{International Journal of Fracture}, vol. 178, no. 1-2, pp. 85--95, Nov.
  2012.

\bibitem{miehe_phase_2016}
C.~Miehe and S.~Mauthe, ``Phase field modeling of fracture in multi-physics
  problems. {{Part III}}. {{Crack}} driving forces in hydro-poro-elasticity and
  hydraulic fracturing of fluid-saturated porous media,'' \emph{Computer
  Methods in Applied Mechanics and Engineering}, vol. 304, pp. 619--655, Jun.
  2016.

\bibitem{alessi_gradient_2015}
R.~Alessi, J.-J. Marigo, and S.~Vidoli, ``Gradient damage models coupled with
  plasticity: {{Variational}} formulation and main properties,''
  \emph{Mechanics of Materials}, vol.~80, pp. 351--367, Jan. 2015.

\bibitem{raina_phasefield_2016}
A.~Raina and C.~Miehe, ``A phase-field model for fracture in biological
  tissues,'' \emph{Biomechanics and Modeling in Mechanobiology}, vol.~15,
  no.~3, pp. 479--496, Jun. 2016.

\bibitem{suh_opensource_2019}
H.~S. Suh and W.~Sun, ``An {{Open-Source FEniCS Implementation Of A Phase Field
  Fracture Model For Micropolar Continua}},'' \emph{International Journal for
  Multiscale Computational Engineering}, vol.~17, no.~6, pp. 639--663, 2019.

\bibitem{wang_inference_2021-1}
\BIBentryALTinterwordspacing
Z.~Wang, J.~Estrada, E.~Arruda, and K.~Garikipati,
  ``\BIBforeignlanguage{en}{Inference of deformation mechanisms and
  constitutive response of soft material surrogates of biological tissue by
  full-field characterization and data-driven variational system
  identification},'' \emph{\BIBforeignlanguage{en}{Journal of the Mechanics and
  Physics of Solids}}, vol. 153, p. 104474, Aug. 2021. [Online]. Available:
  \url{https://linkinghub.elsevier.com/retrieve/pii/S0022509621001459}
\BIBentrySTDinterwordspacing

\end{thebibliography}
\end{document}